\documentclass{PoS}
\rightline{\parbox{4cm}{\large\rm
Edinburgh 2008/49\\
}}

\usepackage{amsmath}
\usepackage{amssymb}
\usepackage{graphicx}
\usepackage{wrapfig}
\usepackage{psfrag}
\newcommand{\be}{\begin{equation}}
\newcommand{\ee}{\end{equation}}
\newcommand{\bc}{\begin{center}}
\newcommand{\ec}{\end{center}}

\title{Investigations of hadron structure on the lattice}

\ShortTitle{Investigations of hadron structure on the lattice}

\author{\speaker{James M. Zanotti}\\
        School of Physics and Astronomy, University of Edinburgh,
        Edinburgh EH9 3JZ, UK\\
        E-mail: \email{jzanotti@ph.ed.ac.uk}}

      \abstract{Lattice simulations of hadronic structure are now
        reaching a level where they are able to not only complement,
        but also provide guidance to current and forthcoming
        experimental programmes at, e.g. Jefferson Lab, COMPASS/CERN
        and FAIR/GSI. 
        In this talk I review the progress that has been made in this
        exciting area in the past year and discuss the advances that
        we can expect to see in the coming year.
        Topics to be covered include form factors (including
        transition form factors), moments of ordinary parton and
        generalised parton distribution functions, moments of
        distribution amplitudes, and magnetic and electric
        polarisabilities.
        I will also highlight the progress being made in determining
        disconnected contributions to hadronic properties.
        Of particular interest here is the size of the contribution to
        various nucleonic properties coming from strange quarks.
      }

\FullConference{The XXVI International Symposium on Lattice Field Theory\\
		 July 14-19 2008\\
		 Williamsburg, Virginia, USA}

\begin{document}

%%%%%%%%%%%%%%%%%%%%%%%%%%%%%%%%%%%%%%%%%%%%%%%%%%%%%%%%%%%%%%%%%%%
%
\section{Introduction}
\label{sec:intro}
%
%%%%%%%%%%%%%%%%%%%%%%%%%%%%%%%%%%%%%%%%%%%%%%%%%%%%%%%%%%%%%%%%%%%

Much of our knowledge about hadronic structure in terms of quark and
gluon degrees of freedom has been obtained from high energy scattering
experiments.
However, as discussed in the talk of Vanderhaeghen
\cite{vanderhaeghen}, there are still many unresolved issues in
hadronic physics that need to be addressed, from both an experimental
and theoretical perspective.
This is one of the main motivations of the 12~GeV Jefferson Lab
upgrade which aims to \cite{JLab}: search for exotic mesons;
study the role of hidden flavours in the nucleon; map out the spin and
flavour dependence of the valence parton distribution functions;
explore nuclear medium effecs; and measure the generalised parton
distribution functions of the nucleon. 
It is imperative that these and other exciting experimental efforts,
such as those at COMPASS/CERN and FAIR/GSI, are matched by modern
lattice simulations which, thanks to recent innovative computer and
algorithmic improvements \cite{ishikawa}, are now capable of reaching
light quark masses ($m_\pi<300$~MeV) and large volumes (>3~fm)
\cite{jansen}.

In this talk I will report on progress made in the past year (for
reviews of results reported in the previous two conferences, see
earlier reviews by Orginos \cite{Orginos:2006zz} and H\"agler
\cite{Hagler:2007hu}) in lattice calculations of many different aspects of
hadronic physics such as the electromagnetic form factors of
$N,\,\pi,\,\rho,\,\Delta$ and transitions in Sec.~\ref{sec:ff},
moments of structure functions in Sec.~\ref{sec:SF}, axial coupling
constants of baryons in Sec.~\ref{sec:axial}, moments of generalised
parton distributions (Sec.~\ref{sec:gpds}) and distribution amplitudes
(Sec.~\ref{sec:da}), disconnected contributions in
Sec.~\ref{sec:disc}, polarisabilities in Sec.~\ref{sec:bff}, and
finally in Sec.~\ref{sec:fin} I summarise the current status of these
topics and point out unresolved issues and directions for the future.

%%%%%%%%%%%%%%%%%%%%%%%%%%%%%%%%%%%%%%%%%%%%%%%%%%%%%%%%%%%%%%%%%%%
%
\section{Electromagnetic Form Factors}
\label{sec:ff}
%
%%%%%%%%%%%%%%%%%%%%%%%%%%%%%%%%%%%%%%%%%%%%%%%%%%%%%%%%%%%%%%%%%%%

The study of the electromagnetic properties of hadrons provides
important insights into the non-perturbative structure of QCD.
The EM form factors reveal important information on the internal
structure of hadrons including their size, charge distribution and
magnetisation.

A lattice calculation of the $q^2$-dependence of hadronic
electromagnetic form factors can not only allow for a comparison with
experiment, but also help in the understanding of the asymptotic
behaviour of these form factors, which is predicted from perturbative
QCD.
Such a lattice calculation would also allow for the extraction of
other phenomenologically interesting quantities such as charge radii
and magnetic moments.
For a recent review see \cite{Arrington:2006zm}.

\subsection{Nucleon Form Factors}
\label{sec:nff}

Phenomenological interest in the electromagnetic form factors of the
proton has been revived by recent Jefferson Lab polarisation
experiments \cite{JLabFF} measuring the ratio of the proton electric
to magnetic, $\mu^{(p)}G_e^{(p)}(q^2)/G_m^{(p)}(q^2)$, and Pauli to
Dirac, $F_2(q^2)/F_1(q^2)$, form factors.
Based on perturbative QCD \cite{Brodsky:1974vy}, the asymptotic
scaling behaviour of these ratios should be independent of $q^2$
(for $G_e/G_m$) or scale as $1/q^2$ (for $F_2/F_1$), however
these experiments showed that $G_e/G_m$ decreases almost linearly with
increasing $q^2$, while $F_2/F_1$ scales as $1/\sqrt{q^2}$.
Additionally, fits of proton and neutron data using
phenomenologically motivated ans\"aze provide for the possibility of a
zero crossing in the isovector electric form factor, $G_e^v$, around
$Q^2 \sim 4.5$\,(GeV/c)$^2$ \cite{Kelly:2004hm}.

The electromagnetic form factors of the neutron are also receiving
plenty of interest at the moment since we know that it has charge
zero, but how is its internal charge distributed and does it have a
positively or negatively charged core \cite{Miller:2008jc}?  
Lattice calculations can provide insights into this distribution since
lattice simulations of three-point functions are performed at the
quark level, and hence they have an advantage over experiment in that
they can directly measure the individual quark contributions to the
nucleon form factors.

On the lattice, we determine the form factors $F_1(q^2)$ and
$F_2(q^2)$ by calculating the following matrix element of the
electromagnetic current
\be
\langle p',\,s'| j^{\mu}(\vec{q})|p,\,s\rangle
\, = 
 \bar{u}(p',\,s')
 \left[ \gamma^\mu F_1(q^2) +
       i\sigma^{\mu\nu}\frac{q_\nu}{2m_N}F_2(q^2) \right] 
 u(p,\,s) \, ,
\label{eq:em-me}
\ee
where $u(p,\,s)$ is a Dirac spinor with momentum, $p$, and spin
polarisation, $s$, $q = p' - p$ is the momentum transfer, $m_N$ is
the nucleon mass and $j_\mu$ is the electromagnetic current.
The Dirac $(F_1)$ and Pauli $(F_2)$ form factors of the proton are
obtained by using $j_\mu^{(p)} = \frac{2}{3}\bar{u}\gamma_\mu u -
\frac{1}{3}\bar{d}\gamma_\mu d$, while for isovector form factors
$j_\mu^v = \bar{u}\gamma_\mu u - \bar{d}\gamma_\mu d$.
It is common to rewrite the form factors $F_1$ and $F_2$ in terms of
the 
electric and magnetic Sachs form factors, 
$G_e= F_1 + q^2/(2m_N)^2\, F_2$ and $G_m= F_1 + F_2$.

If one is using a conserved current, then (e.g. for the proton)
$F_1^{(p)}(0) = G_e^{(p)}(0) =1$ gives the electric charge,
while $G_m^{(p)}(0) = \mu^{(p)} = 1 + \kappa^{(p)}$
gives the magnetic moment, where $F_2^{(p)}(0) = \kappa^{(p)}$ is the
anomalous magnetic moment.
From Eq.~(\ref{eq:em-me}) with see that $F_2$ always appears with a
factor of $q$, so it is not possible to extract a value for $F_2$ at
$q^2=0$ directly from our lattice simulations.
Hence we are required to extrapolate the results we obtain at finite
$q^2$ to $q^2=0$.
Form factor radii, $r_i=\sqrt{\langle r_i^2\rangle}$, are defined as
the slope of the form factor at $q^2=0$.

\begin{figure}[t]
     \vspace*{-7mm}
   \begin{minipage}{0.48\textwidth}
      \centering
          \includegraphics[clip=true,width=0.9\textwidth]{Figures/mfr_1_c.eps}
\caption{Comparison of results for the isovector Dirac radius, $r_1$.}
\label{fig:RBC-f1-r1}
     \end{minipage}
     \hspace{0.1cm}
    \begin{minipage}{0.48\textwidth}
      \centering
     \vspace*{-7mm}
          \includegraphics[width=0.93\textwidth]{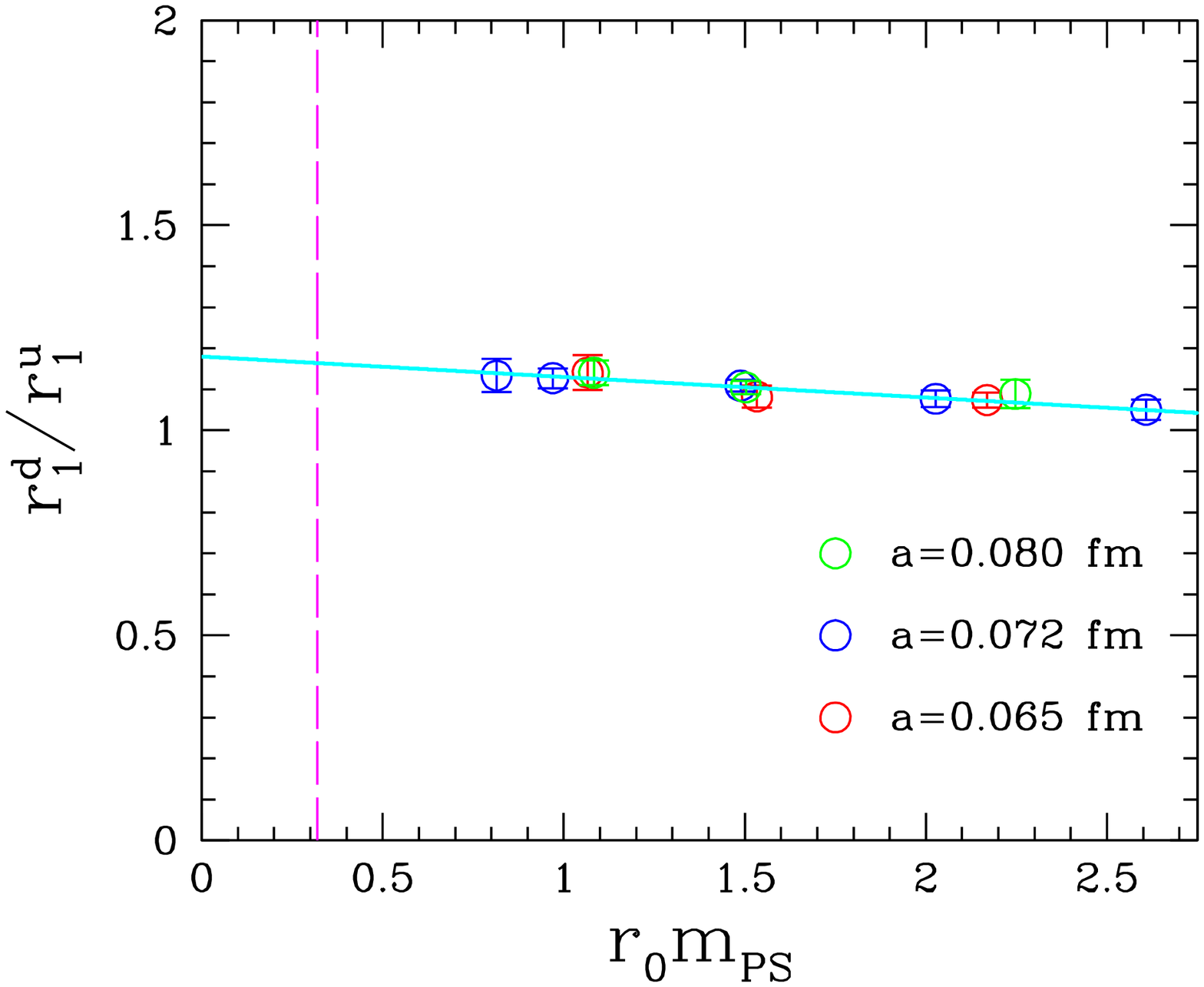}
\caption{Ratio of Dirac radii for $u$ and $d$-quarks from
  QCDSF. Dashed line indicates physical $m_\pi$.}
\label{fig:QCDSF-rr12du}
     \end{minipage}
 \end{figure}

In Fig.~\ref{fig:RBC-f1-r1} we see results for the isovector Dirac
radius from several different fermion actions.
RBC/UKQCD presented an update from the $N_f=2+1$ domain wall fermion
run on $24^3\times 64$ lattices with $a^{-1}=1.729$\,GeV
\cite{Ohta:2008kd} (red circles), while LHPC updated their mixed
action (DWF valence, asqtad sea) results at their lightest pion masses
\cite{Bratt:2008uf} (green right triangles).
Additionally, LHPC have started running on the $32^3\times 64$ DWF
configurations with $a^{-1}\approx 2.4$\,GeV generated by the RBC/UKQCD
collaborations, and preliminary results from these runs are shown by
the black upside-down triangles \cite{Bratt:2008uf}.

These latest results are compared with earlier quenched and $N_f=2$
Wilson \cite{Gockeler:2003ay,Alexandrou:2006ru} and DWF
\cite{Sasaki:2007gw,Lin:2008uz} results.
We observe agreement between the different lattice
formulations, while any discrepancies are an indication for systematic
uncertainties, such as finite volume effects, discretisation errors,
etc.
The overall pattern is typical of lattice results for
$r_1$, i.e. the lattice results lie below experiment with little
variation as a function of $m_\pi^2$.
Investigations using chiral perturbation theory predict that these
radii should increase dramatically close to the chiral limit
\cite{Gockeler:2003ay,Young:2004tb}.
Current results indicate that in order to see such curvature, one needs to
perform simulations at $m_\pi<300$~MeV.

During the conference, we also saw a preliminary analysis from the
European Twisted Mass Collaboration using $N_f=2$ twisted mass
fermions with pion masses down to $m_\pi\simeq 313$~MeV at a single
lattice spacing, $a=0.089(1)$\,fm \cite{Korzec}, and results are
forthcoming.

Finally, QCDSF have been studing the $q^2$-dependence of the
individual quark contributions to the nucleon form factors.
In Fig.~\ref{fig:QCDSF-rr12du} we see some results for the ratio of
the $d$- to $u$-quark contributions to the proton's Dirac radius.
Here we clearly see that $r_{1}^d>r_{1}^u$ for all simulated quark
masses (the same behaviour is seen for $r_2$), indicating that the
$d\,(u)$-quarks are more broadly distributed than $u\,(d)$-quarks in
the proton (neutron).
Note that disconnected contributions were not considered in this
study.

\subsection{Accessing small $Q^2$: Partially twisted boundary conditions}
\label{sec:tbc}

On a lattice of spatial size, $L$, momenta are discretised in units of
$2\pi/L$. 
Modifying the boundary conditions of the valence quarks
\cite{Sachrajda:2004mi}
$\psi(x_k+L)=e^{i\theta_k}\psi(x_k),\ (k=1,2,3)$
allows one to tune the momenta continuously $\vec{p} +
\vec{\theta}/L$.
Momentum transfer in a matrix element between states with initial and
final momenta, $\vec{p}_i +\vec{\theta}_i/L$ and $\vec{p}_f
+\vec{\theta}_f/L$, respectively, then reads
$
q^2=(p_f-p_i)^2=\Big\{[E_f(\vec p_f, \vec{\theta}_f)-E_i(\vec
  p_i,\vec{\theta}_i)]^2
        -\big[(\vec p_f+\vec{\theta}_f/L)
          -(\vec p_i+\vec{\theta}_i/L)\big]^2\Big\}\ ,
$
where $E(\vec p, \vec{\theta})=\sqrt{m^2+(\vec
  p+\vec{\theta}/L)^2}$.

\begin{figure}[t]
     \vspace*{-4mm}
   \begin{minipage}{0.48\textwidth}
      \centering
     \vspace*{1mm}
\includegraphics[clip=true,width=1.03\textwidth]{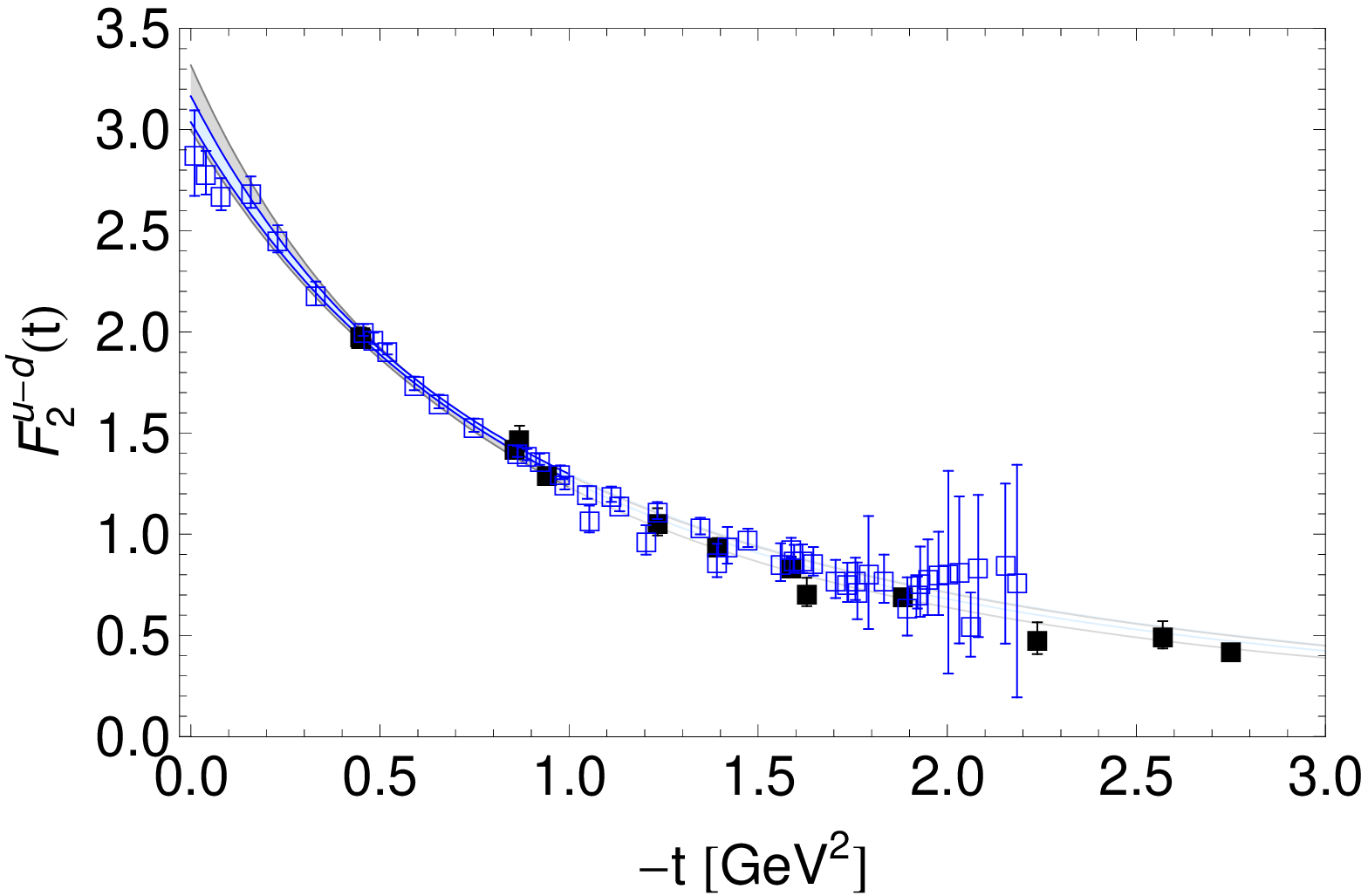}
\caption{Pauli form factor, $F_2(q^2)$ together with an
  extrapolation to $q^2=0$. Open blue symbols
  indicate results using partially twisted bc's.}
\label{fig:ff-tbc}
     \end{minipage}
     \hspace{0.1cm}
    \begin{minipage}{0.48\textwidth}
      \centering
\includegraphics[width=1.0\textwidth]{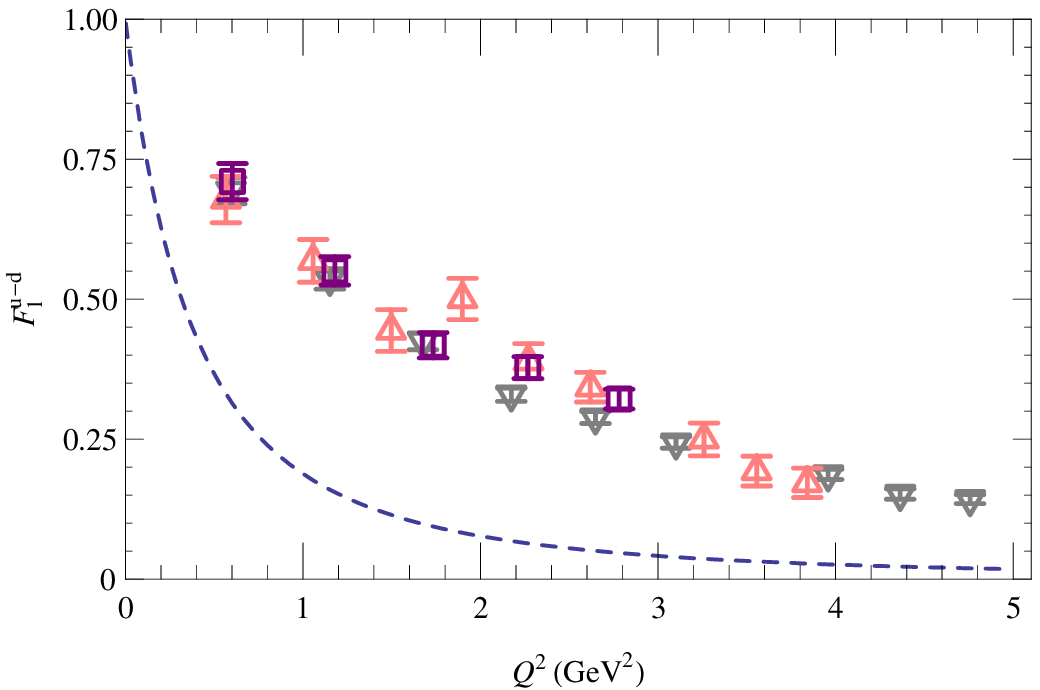}
\caption{$F_1(Q^2)$ from \cite{HW-latt} obtained using a variational
  analysis for $m_\pi\simeq 480$ (pink), 720 (purple), 1100 (grey)
  MeV.  Dashed curve indicates a fit to experimental data.}
\label{fig:JLab-largeQ2}
     \end{minipage}
 \end{figure}

$F_2$ is particularly interesting since it cannot be measured directly
at $q^2=0$ to obtain magnetic moments.
Hence it needs to be extrapolatd from finite $q^2$ which can not only
increase the error, but can also introduce a model dependence into
the result.
As can be seen in Fig.~\ref{fig:ff-tbc} from the QCDSF collaboration
\cite{Phil-latt}, results obtained by using partially twisted bc's
(open blue symbols) help to constrain the extrapolation to $q^2=0$.

Twisted boundary conditions, however, introduce additional finite
volume (FV) effects $\sim e^{-m_\pi L}$, which were shown
to be small
for the pion form factor in the Breit frame \cite{Jiang:2008te}, but
can be substantial for isovector nucleon form factors
\cite{Tiburzi:2006px}.
In \cite{Jiang:2008ja} it was shown that when tbcs are applied only to
the active quarks attached to the current, the FV corrections depend
on an unphysical and unknown parameter.
They also found that the FV corrections are largest for the magnetic
form factor with small twists.
Indeed, the partially twisted bc results at small $q^2$ in
Fig.~\ref{fig:ff-tbc} appear to be suppressed compared to the overall
fit, which is the expectation from
\cite{Tiburzi:2006px,Jiang:2008ja}. 

\subsection{Large $Q^2$}
\label{sec:largeq}

Lattice calculations suffer from noise at large $Q^2$. 
This typically restricts the range of available momentum transfers to
$Q^2<4\,{\rm GeV}^2$.
Earlier attempts by the LHP Collaboration to access the
electromagnetic form factors out to $Q^2\sim 6$~GeV$^2$ in the Breit
frame, $(\vec{p}\,' = -\vec{p} = 2\pi \vec{n}/L)$
\cite{Edwards:2005kw} revealed that the relative error in
$F_1^{v}(Q^2)$, at fixed pion mass increases as $n^4$.
They found that in order to achieve a point at
$Q^2\approx 6$~GeV$^2$ with a relative error of 30\%, they would
have to increase the statistical accuracy by at least a factor of 50.
Furthermore, to compound the difficulty, it was observed that the
relative error in the isovector Dirac form factor increased with
$\sim1/m^4_\pi$.  

This has led the JLab group to attempt a study of these form factors
using variational methods \cite{HW-latt}.
Their initial quenched study is performed  on a $16^3\times 64$
anisotropic lattice ($\xi=3$) using the Wilson gauge action and clover
fermion action with 3 quark masses corresponding to pion masses of
1100, 720 and 480 MeV.

To extract (\ref{eq:em-me}) from a lattice 3-point function, accurate
knowledge on the overlap factors and masses is required.
Usually these are cancelled by constructing a ratio of 3pt and 2pt
functions, however the drawback here is that often one needs to use a
2pt function with large momentum at large Euclidean times, which
introduces additional statistical noise.
An additional problem could arise if a smeared source is used that has
been tuned at $\vec{p}=0$, but may not be ideal at large $\vec{p}$.
To circumvent these issues, the JLab group use three different choices
of gaussian smearing and then solve a generalised eigenvalue problem
to obtain the overlap factors and masses from the two point functions.
These are then used in the 3pt correlator to solve for the form
factors.

Their preliminary result for $F_1(Q^2)$ using this method is shown in
Fig.~\ref{fig:JLab-largeQ2}.
Encouragingly, we see that they are able to find a clean signal up to
$Q^2\approx 5$\,GeV$^2$.
Simulations with dynamical fermions and lighter quark masses are
now starting.

\subsection{Nucleon-$P_{11}$ (Roper) Form Factors}
\label{sec:nroper}

Using the methods outline in the previous section, the JLab group have
performed a quenched study of the transition form factors of the
ground-state nucleon to its $P_{11}$ excited state \cite{Lin:2008qv}
\be
\langle N_2|j_\mu(\vec{q})|N_1\rangle = \bar{u}_{N_2}(p')\bigg[
F_1(q^2)\bigg( \gamma_\mu - \frac{q_\mu}{q^2}q\!\!\!\!\!\!\!\not\ \bigg) +
\sigma_{\mu\nu}q_\nu\frac{F_2(q^2)}{M_{N_1}+M_{N_2}}\bigg] u_{N_1}(p)\ ,
\ee
which are extracted from the correlators
using a variational analysis.

\begin{wrapfigure}{r}{0.45\textwidth}
\bc
\hspace*{-9mm}
\includegraphics[clip=true,width=0.55\textwidth]{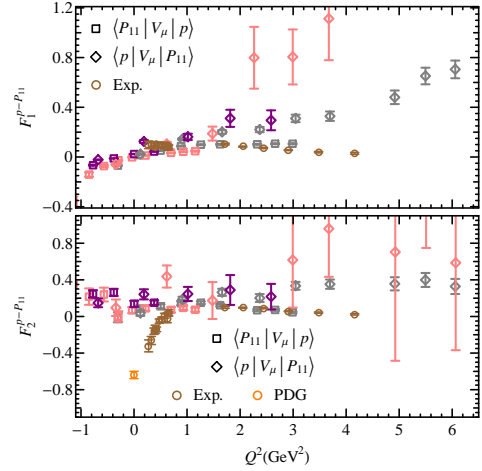}
\caption{Proton-$P_{11}$ 
  factors from a quenched study \cite{Lin:2008qv}. Masses as in
  Fig.~{\protect\ref{fig:JLab-largeQ2}}.}
\label{fig:N-Roper}
\vspace*{-7mm}
\ec
\end{wrapfigure}

Results for $\langle P_{11}|V_\mu|p\rangle$ from this quenched initial
study are shown in Fig.~\ref{fig:N-Roper}, where we clearly see that
it is possible to obtain a signal in such channels.
While the behaviour of the results is different to that of the
experimental data, this is probably a result of the heavy quark
masses currently being used.
Investigations are now under way with dynamical fermions and lighter
quark masses.

\subsection{Pion Form Factor}
\label{sec:pionff}

The pion form factor, $F_\pi(Q^2)$, has recently received a surge of
interest from several lattice groups.
This is partly due to the fact that the pion is the easiest
hadron to study on the lattice, making it the perfect candidate for
testing new techniques (e.g. twisted boundary conditions, all-to-all
propagators).
In addition to this, $F_\pi(Q^2)$ is an interesting
quantity to study phenomenologically since its asymptotic
($Q^2\to\infty$) normalisation is known from $\pi\to\mu +\nu$ decay,
and hence it allows us to study the transistion from the soft to hard
regimes.
At low $Q^2$ the $F_\pi(Q^2)$ is measured directly by scattering high
energy pions from atomic elections \cite{Amendolia:1986wj}, however
measurements at high $Q^2$ require quasi-elastic scattering off
virtual pions \cite{Brauel:1979zk} which leads to a model dependence
in the extraction of the form factor from experimental data; a source
of systematic error not present in a lattice calculation.

\begin{figure}[t]
   \begin{minipage}{0.48\textwidth}
      \centering
\vspace*{-7mm}
\psfrag{xlabel}[ct][b][1][180]{\large $Q^2[{\rm GeV}^2]$}
\psfrag{ylabel}[cb][t][1][0]{\large $F_{\pi}(Q^2)$}
\includegraphics[height=0.95\textwidth,angle=-90]{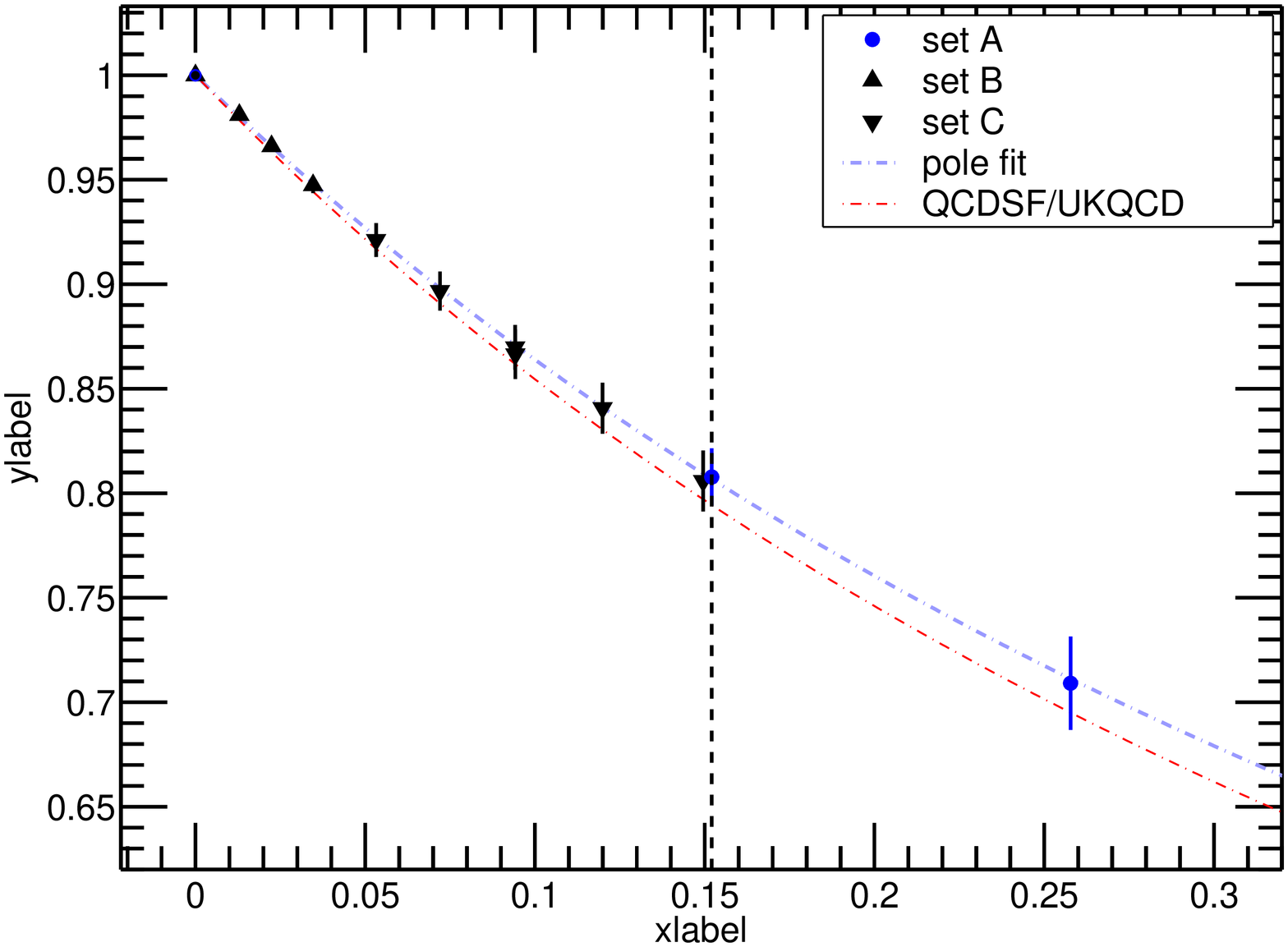}
\caption{$F_\pi(Q^2)$ determined using stochastic propagators
  (one-end trick) with partially twisted bc's 
using DWF 
with $m_\pi\approx 330$\,MeV.}
\label{fig:fpiUKQCD}
     \end{minipage}
     \hspace{0.1cm}
    \begin{minipage}{0.48\textwidth}
      \centering
\includegraphics[width=0.95\textwidth]{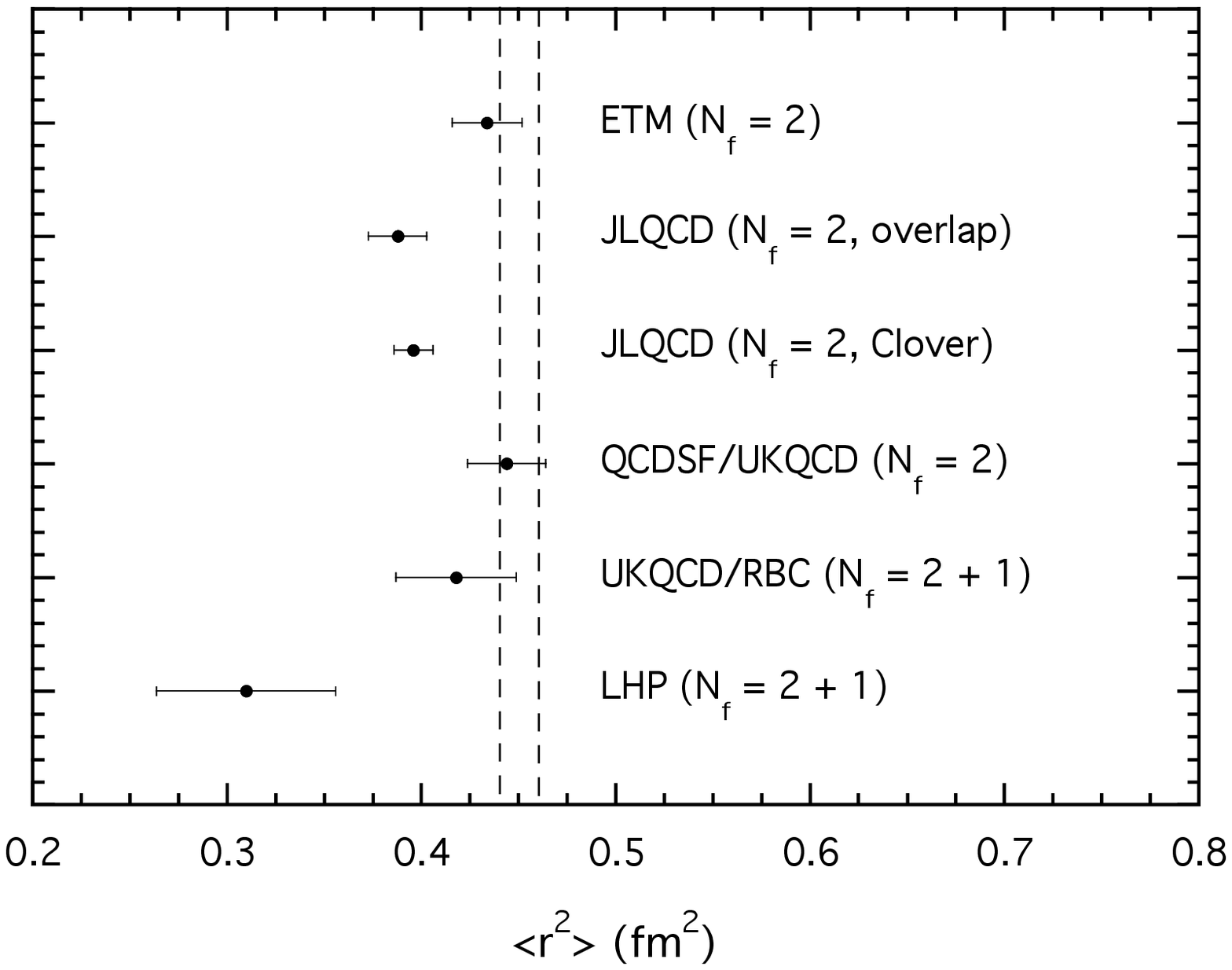}
\caption{Comparison of latest lattice results of $\langle r_\pi^2\rangle$
with the experimental value \cite{Yao:2006px} (dashed lines).}
\label{fig:fpicompare}
     \end{minipage}
 \end{figure}

Recently, RBC/UKQCD have used stochastic propagators with a single
spin/colour source calculated using the so-called ``one-end-trick''
\cite{Foster:1998vw}, together with twisted boundary conditions
\cite{Sachrajda:2004mi} to calculate the $F_\pi(Q^2)$ at small values
of the momentum transfer \cite{Boyle:2008yd}.
The results from this study are presented in Fig.~\ref{fig:fpiUKQCD}.
Here the smallest momentum transfer available on this lattice using
periodic bc's is denoted by the vertical dashed line and the
results for $F_\pi(Q^2)$ for these by filled circles.
The results obtained using twisted boundary conditions are given by
the triangles, and we clearly see that they smoothly fill the gap
between the first fourier momentum and $Q^2=0$.

Using their results at the smallest 3 values of $Q^2$, the authors
computed the pion charge radius and, using the NLO expression from
ChPT \cite{Gasser:1983yg}
\be
\langle r_\pi^2\rangle_{\mathrm{SU}(2),\mathrm{NLO}} =
 -\frac{12l_6^r}{f^2} - \frac1{8\pi^2f^2}
 \Big(\log\frac{m_\pi^2}{\mu^2}+1\Big)\ ,
\ee
and the value of the pion decay constant in the chiral limit
\cite{Allton:2008pn}, are able to determine the LEC,
$l_6^r(m_\rho)=-0.0093(10)$.
Evaluting the expression using the physical $m_\pi$ gives $\langle
r_\pi^2\rangle=0.418(28)$\,fm$^2$, compared with $\langle
r_\pi^2\rangle_{\rm exp}=0.452(11)$\,fm$^2$.

The ETM Collaboration are also using the combination of stochastic
propagators and twisted bc's to determine the pion form factor (and
subsequently, $\langle r_\pi^2\rangle$) on their $N_f=2$ twisted mass
configurations \cite{Simula:2007fa}.
They have calculated $\langle r_\pi^2\rangle$ at a number of pion
masses and extrapolated their values to the physical pion mass using
2-loop ChPT \cite{Bijnens:1998fm}, obtaining $\langle
r_\pi^2\rangle=0.396(10)$\,fm$^2$.

We also saw an updated analysis from the JLQCD collaboration with
doubled statistics \cite{JLQCD:2008kx}.
They are using all-to-all propagators to calculate $F_\pi(Q^2)$ with
$N_f=2$ overlap fermions on a $16^3\times 32$ lattice with light quark
masses down to a sixth of the strange quark mass.
By comparing NLO and NNLO ChPt, they find the two-loop contribution
to $\langle r_\pi^2\rangle$ to be significant in the simulated region.
Hence they performed a joint two-loop fit to $\langle r_\pi^2\rangle$
and $\langle r_{\pi,S}^2\rangle$ (pion scalar form factor, see
Sec.~\ref{sec:scalarff}), from which they obtain $\langle
r_\pi^2\rangle=0.404(22)(22)$\,fm$^2$.

In Fig.~\ref{fig:fpicompare}, we present the current status of lattice
determinations of $\langle r_\pi^2\rangle$ by comparing the latest
results with earlier determinations \cite{Brommel:2006ww} and the
experimental value \cite{Yao:2006px}.
While there is a slight scatter, the general trend of the lattice
results, even after attempts at including chiral logs, is to lie low
compared to the experimental value.
Whether this can be explained by finite volume effects, discretisation
errors, or even the application of ChPT at such large quark masses,
will require further investigation.

\subsection{$\Delta$ Electromagnetic Form Factors}
\label{sec:decff}

The matrix element of the electromagnetic current between spin-3/2 states
has the form (c.f. Eq.~(\ref{eq:em-me}) for spin-1/2 states)
\be
\langle \Delta(\vec{p}\,',s')| j^\mu |\Delta(\vec{p},s)\rangle = 
\sqrt{\frac{m_B^2}
{E_B(\vec{p}\,') E_B(\vec{p})}}
\bar{u}_\sigma (\vec{p}\,',s')O^{\sigma\mu\tau}u_\tau(\vec{p},s)\ ,
\ee
where $u_\sigma(p,s)$ is a Rarita-Schwinger spin-vector, $M_B$ is the
mass of the decuplet baryon and
\be 
O^{\sigma\mu\tau} = -g^{\sigma\tau} \left[ a_1(q^2) \gamma^\mu +
  \frac{a_2(q^2)}{2m_B} (p'^\mu + p^\mu)\right] - 
\frac{q^\sigma q^\tau}{4m_B^2} 
  \left[ c_1(q^2) \gamma^\mu +
    \frac{c_2(q^2)}{2m_B} (p'^\mu + p^\mu)\right]\ .  
\ee
The parameters $a_1,\ a_2,\ c_1$ and $c_2$ are independent covariant
vertex function coefficients.
For decuplet baryons, there are four multipole form factors, $G_{E0},\
G_{E2},\ G_{M1},\ G_{M3}$, which are defined in terms of $a_1,\,
a_2,\, c_1,\, c_2$, and are referred to as charge $(E0)$,
electric-quadrupole $(E2)$, magnetic-dipole $(M1)$ and
magnetic-octupole $(M3)$ form factors, respectively.

While the $E0$ and $M1$ form factors give access to charge radii and
magnetic moments in the same way as for spin-1/2 baryons, the $E2$ and
$M3$ moments accessible in spin-3/2 systems provide insights into the
shape of decuplet baryons and have the potential to discriminate
between various model descriptions of hadronic phenomena.

\begin{figure}[t]
   \begin{minipage}{0.48\textwidth}
      \centering
\includegraphics[angle=90,width=0.9\textwidth]{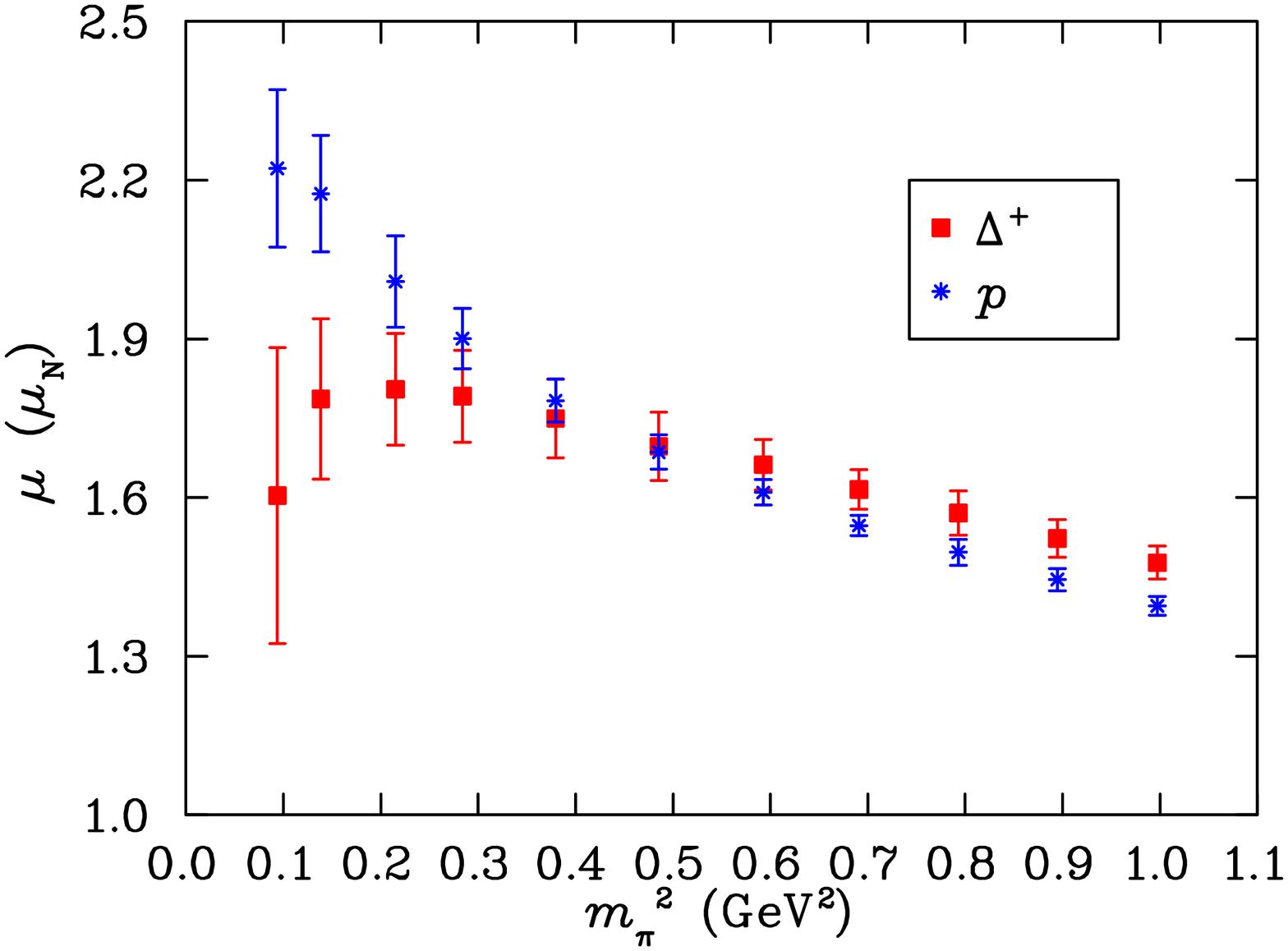}
\caption{Quenched Adelaide results for magnetic moments of the proton
  and $\Delta^+$.}
\label{fig:CSSMdec}
     \end{minipage}
     \hspace{0.1cm}
    \begin{minipage}{0.48\textwidth}
      \centering
\includegraphics[clip=true,width=0.9\textwidth]{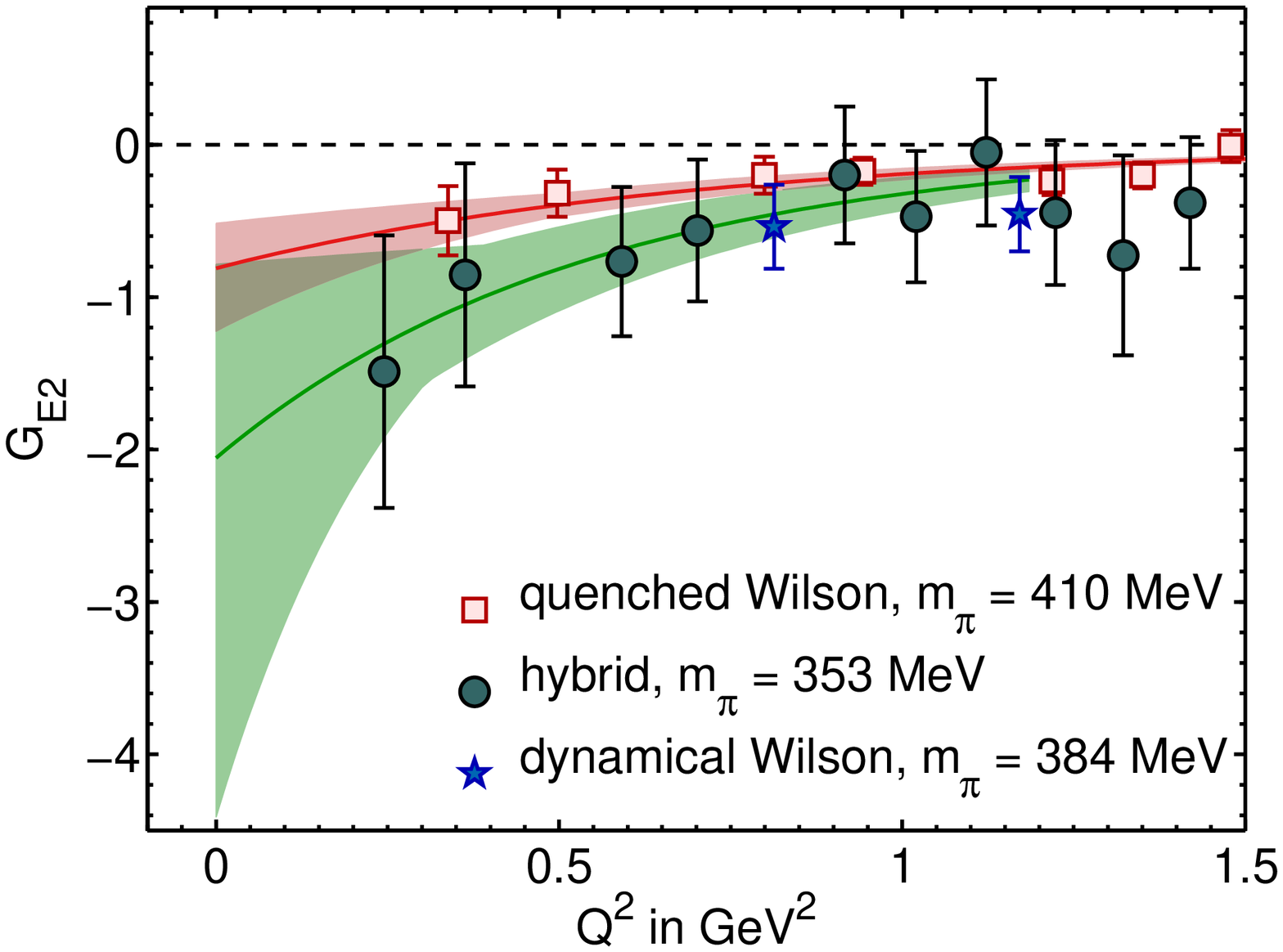}
\vspace*{-1mm}
\caption{$G_{E2}$ form factors from the Cyprus
  group \cite{Alexandrou:2008bn} with exponential fits.}
\label{fig:CyprusDec}
     \end{minipage}
 \end{figure}

The Adelaide group are in the process of finalising their analysis of
the multipole form factors of the full baryon decuplet \cite{CSSMdec}
in the quenched approximation.
Their findings for the magnetic moment of the $\Delta^+$ is shown in
Fig.~\ref{fig:CSSMdec} and the results are compared with earler proton
results \cite{Boinepalli:2006xd}.
A simple quark model predicts that they should be equal, however, due
to differing pion-loop contributions, the proton and $\Delta^+$
magnetic moments are expected to differ at the physical pion mass
\cite{Cloet:2003jm}.
In fact, quenched ChPT predicts that the pion-loop contributions for
the $\Delta^+$ come with an opposite sign to that of full QCD
\cite{Labrenz:1996jy}, and the results in Fig.~\ref{fig:CSSMdec}
confirm this prediction.

To confirm that this is a quenched artifact, it is important to
perform simulations in full QCD at light enough quark masses.
The Cyprus group have started to perform such simulations
\cite{Alexandrou:2008bn} and 
they are currently simulating at pion masses down to $\sim 350$~MeV
and so are now starting to enter the region where the quenched results
``bend down''.

Both groups find that $G_{E2}$ is negative as shown in
Fig.~\ref{fig:CyprusDec} from the Cyprus group, indicating that
$\Delta$ is oblate.
The Adelaide group find that $G_{M3}$ deviates from zero only at small
quark masses, while the Cyprus group only have a result at a single
quark mass where they find that $G_{M3}$ consistent with zero, so it
will be interesting to see if their results will deviate from zero as
their results at smaller quark masses start becoming available.

\subsection{$\rho$ Electromagnetic Form Factors}
\label{sec:rhoff}

The QCDSF collaboration has recently started an investigation into the
electromagnetic form factors of the $\rho$ meson \cite{QCDSFrho}.
The matrix element of the electromagnetic current between spin-1
states is decomposed in terms of three form factors, $G_E(Q^2),\,
G_M(Q^2),\,G_Q(Q^2)$.
Of particular interest is the value of the quadrupole form factor at
zero momentum transfer, $G_Q(Q^2=0)$, which gives the quadrupole
moment; a non-zero value would indicate spatial deformation of the
$\rho$ meson.

\begin{wrapfigure}{r}{0.45\textwidth}
\bc
\vspace*{-8mm}
\includegraphics[angle=90,width=0.45\textwidth]{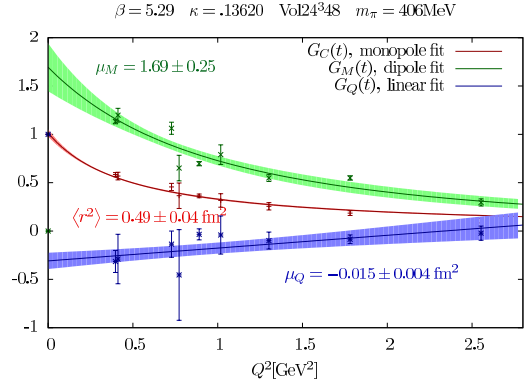}
\vspace*{-2mm}
\caption{$\rho$ form factors from a $24^3\times 48$ lattice with
  $m_\pi=406$\,MeV, $\beta=5.29,\, \kappa=0.13620$}
\label{fig:QCDSFrho}
\vspace*{-4mm}
\ec
\end{wrapfigure}

Fig.~\ref{fig:QCDSFrho} shows the Sachs form factors for the smallest
pion mass analysed $(m_\pi\approx 400\,{\rm MeV})$.
The electric form factor is fitted with a monopole ansatz and from the
slope of the form factor at $Q^2=0$, the charge radius is computed.
After an extrapolation linear in $m_\pi^2$ to the physical pion mass,
they find the preliminary result $\langle r_\rho^2\rangle = 0.49(5)\
{\rm fm}^2$, although it is reasonable to expect some chiral curvature
to enhance this value.

For the magnetic form factor, it is not possible to calculate directly
at $Q^2=0$, which is needed for the determination of the magnetic
moment ($g$-factor).
Hence, the results at $Q^2\ne 0$ are extrapolated to $Q^2=0$ with a
dipole ansatz.
As discussed in Sec.~\ref{sec:tbc}, twisted bc's have the potential to
help here.

Extrapolating linearly in $m_\pi^2$ to the physical pion mass gives
$g_\rho=1.6(1)$, smaller than the quenched result from the Adelaide
group \cite{Hedditch:2007ex} and a study using the background field
method \cite{Lee:2008qf}.

Similar to the magnetic form factor, the quadrupole form factor needs
to be extrapolated from $Q^2\ne 0$ to $Q^2=0$ to obtain the quadrupole
moment. 
The form factor is fitted linearly in $Q^2$, although again twisted
bc's will help to determine if this is a valid assumption, and the
resulting moments linearly in $m_\pi^2$.
The authors find a small negative result $\mu_q=-0.017(2)\ {\rm
  fm}^2$, in agreement with \cite{Hedditch:2007ex}.
The negative result is interpreted as the $\rho$ meson having an
oblate shape, an interpretation enhanced by recent results using
density-density correlators \cite{Alexandrou:2008ru}.

%%%%%%%%%%%%%%%%%%%%%%%%%%%%%%%%%%%%%%%%%%%%%%%%%%%%%%%%%%%%%%%%%%%
%
\subsection{$N\to\Delta$ Transition Form Factors}
\label{sec:ndelta}
%
%%%%%%%%%%%%%%%%%%%%%%%%%%%%%%%%%%%%%%%%%%%%%%%%%%%%%%%%%%%%%%%%%%%

We have seen in Sec.~\ref{sec:decff} that there is emerging evidence
that the quadrupole moment of the $\Delta$ is non-zero, indicating
that the $\Delta$ is not spherically symmetric.
The nucleon, on the other hand, being a spin-1/2 particle, doesn't
have a measurable quadrupole moment, however it still may possess an
intrinsic quadrupole moment and thus also be spatially deformed.
A possible way to search for such non-zero amplitudes is through the
study of spin-1/2 to spin-3/2 ($\gamma N\to\Delta$) transitions, which
are also accessible in lattice simulations
\cite{Leinweber:1992pv,Alexandrou:2004xn,Alexandrou:2007dt}.

\begin{figure}
\bc
\includegraphics[width=7cm]{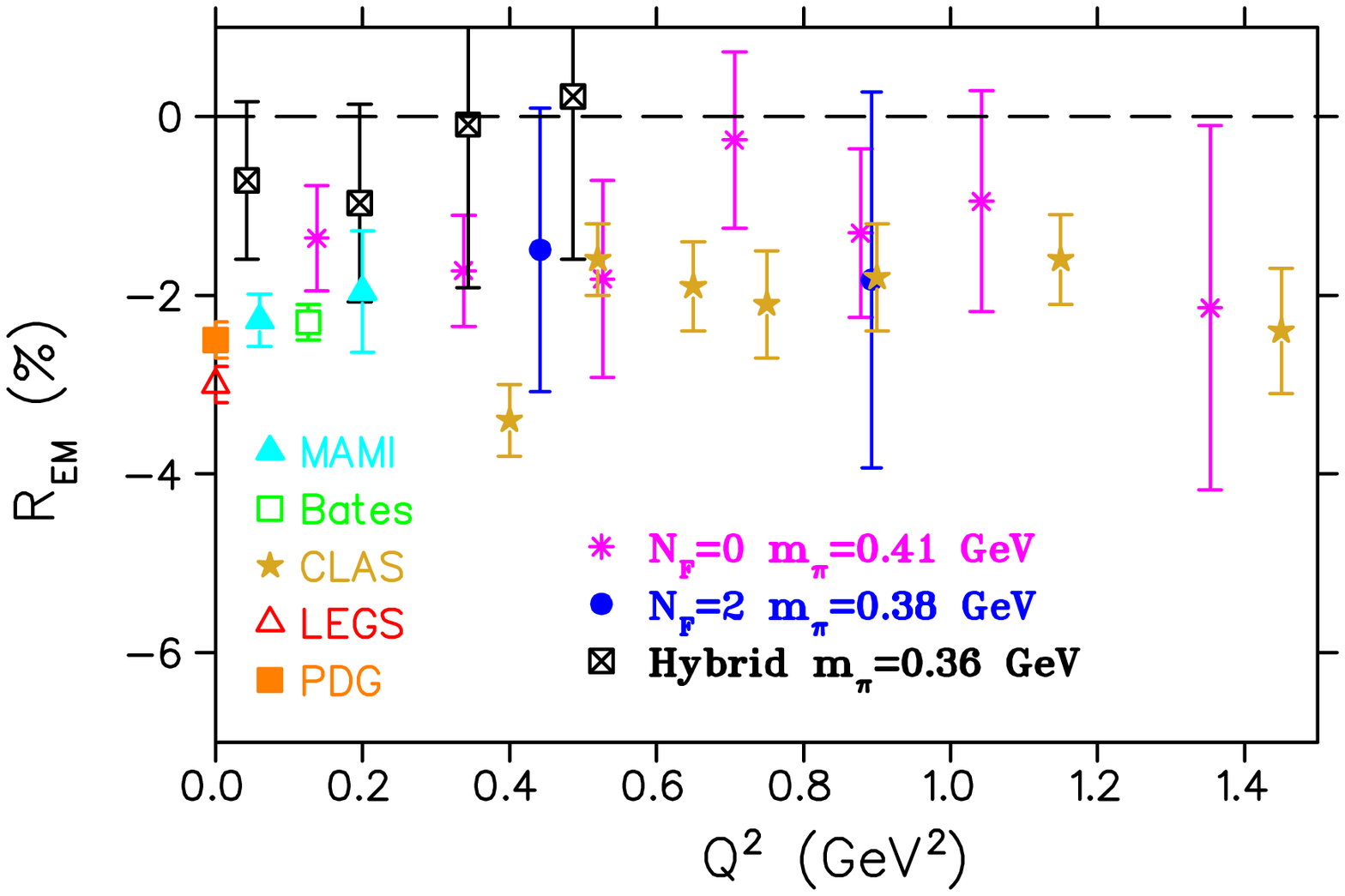}
\includegraphics[width=7cm]{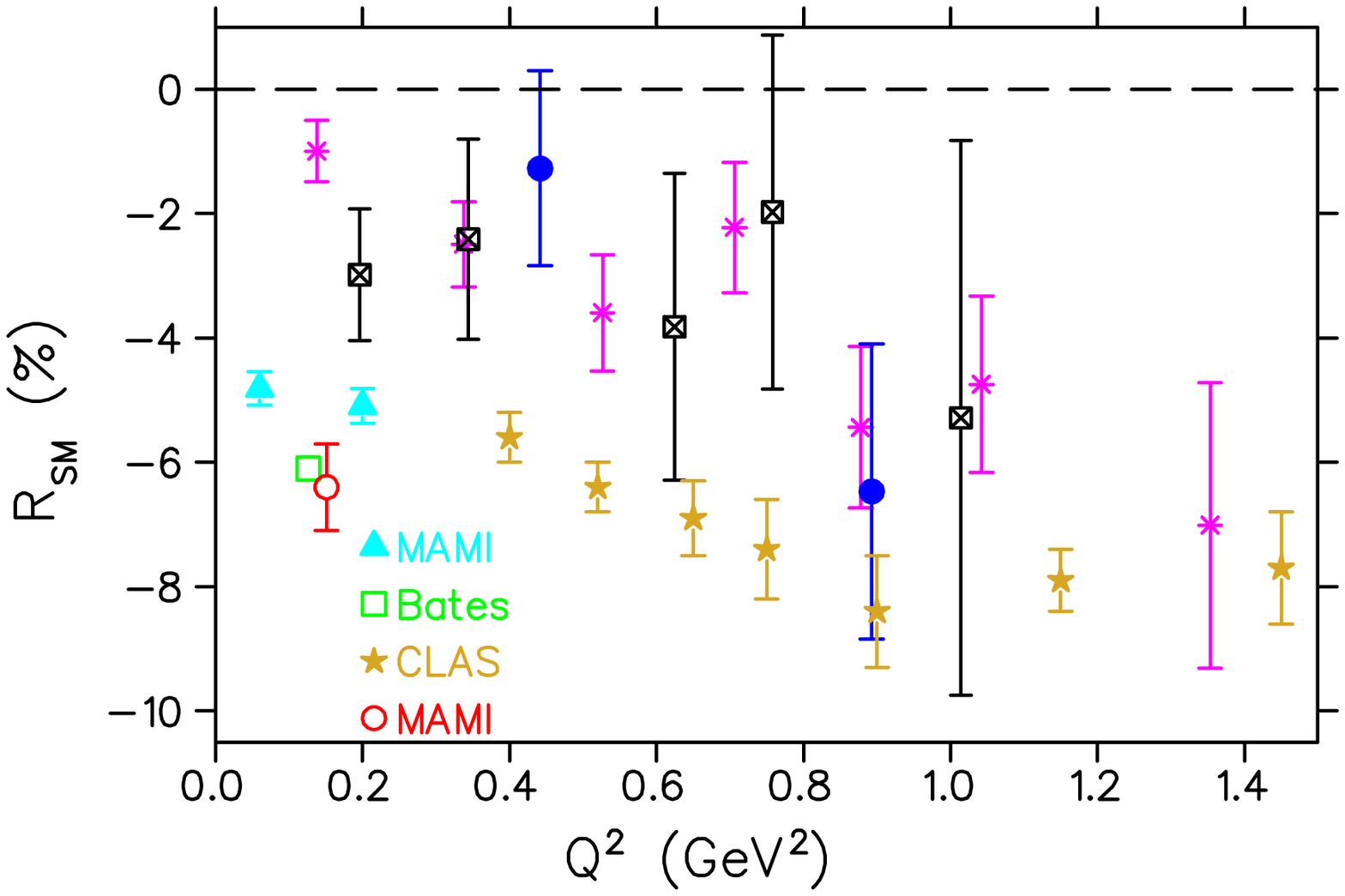}
\caption{$R_{EM}$ and $R_{SM}$ from \cite{Alexandrou:2007dt}. }
\label{fig:vecND}
\ec
\end{figure}

The matrix element for the vector $N\to\Delta$ transition is defined
in terms of three form factors $G_{M1},\,G_{E2},\,G_{C2}$ which are
known as the magnetic dipole, electric quadrupole and Coulomb
quadrupole form factors, respectively.
While the magnetic dipole is dominant, it is possible to search for
non-zero quadrupole form factors by considering the following ratios
measured in the lab frame of the $\Delta$
\be
R_{EM}(EMR) = - \frac{G_{E2}(Q^2)}{G_{M1}(Q^2)}\ ,\quad
R_{SM}(CMR) = -
\frac{|\vec{q}|}{2m_\Delta}\frac{G_{C2}(Q^2)}{G_{M1}(Q^2)}
\label{eq:remrsm}
\ee
Precise experimental data exists for these ratios and strongly suggest
deformation of $N$ and $\Delta$.
This has recently been confirmed in a full QCD simulation by the
Cyprus group \cite{Alexandrou:2007dt}, as we can clearly see in
Fig.~\ref{fig:vecND}.
While $R_{EM}=0$ cannot be ruled out with the current precision on
the hybrid run, $R_{SM}$ is clearly negative, in agreement with
experiment.

%%%%%%%%%%%%%%%%%%%%%%%%%%%%%%%%%%%%%%%%%%%%%%%%%%%%%%%%%%%%%%%%%%%
%
\section{Moments of Structure Functions}
\label{sec:SF}
%
%%%%%%%%%%%%%%%%%%%%%%%%%%%%%%%%%%%%%%%%%%%%%%%%%%%%%%%%%%%%%%%%%%%

\subsection{Nucleon Momentum Fraction, $\langle x\rangle$}
\label{sec:x}

Much of our knowledge about QCD and the structure of the nucleon has
been derived from deep inelastic scattering experiments where cross
sections are determined by its structure functions.
Through the operator product expansion, the first moment of these
structure functions are directly related to the momentum fractions
carried by the quarks and gluons in the nucleon, $\langle
x\rangle_{q,g}$, whose sum must be $\sum_q \langle x\rangle_q +
\langle x\rangle_g =1$.
The scale and scheme dependence of $\langle x\rangle_{q}$ and
$\langle x\rangle_{g}$ cancels out in the sum.

Hence the quark momentum fractions are interesting phenomenologically
and have been studied on the lattice for some time.
In fact, lattice studies of $\langle x\rangle_{q}$ are notorious in
that all lattice results to date at heavy quark masses exhibit an
almost constant behaviour in quark mass towards the chiral limit and
are almost a factor of two larger than phenomenologically accepted
results, e.g. $\langle x\rangle_{u-d}^{\rm MRST}= 0.157(9)$,
leading many a lattice practitioner to scratch their head and wonder
``Will this thing ever bend down?'', as predicted in
\cite{Detmold:2001jb}.

To date, only connected contributions have been simulated to high
precision, hence results are usually quoted for isovector quantities
where disconnected contributions cancel.
For the latest progress on disconnected calculations, see
Sec.~\ref{sec:disc}. 

\begin{wrapfigure}{r}{0.45\textwidth}
\bc
\vspace*{-3mm}
\includegraphics[width=0.43\textwidth]{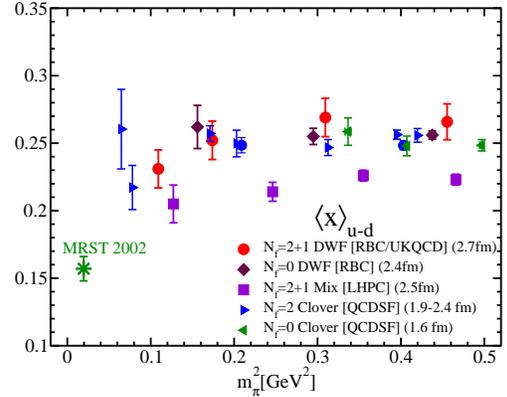}
\caption{$\langle x\rangle_{u-d}$ 
from RBC/UKQCD (DWF), QCDSF (Clover) and LHPC (Mixed) }
\label{fig:RBCx}
\vspace*{-3mm}
\ec
\end{wrapfigure}
Dynamical configurations are now becoming available at quark masses
light enough to enable calculations in the area where such bending is
predicted to set in.
During the conference, RBC/UKQCD presented their findings from their
$N_f=2+1$ DWF configurations with pion masses as low as $m_\pi\approx
330$\,MeV \cite{Ohta:2008kd}.
Results in the $\overline{\rm MS}$ scheme at 2~GeV are shown in
Fig.~\ref{fig:RBCx} and are compared with the latest results from the
QCDSF \cite{Brommel:2008tc} and LHP \cite{Hagler:2007xi}
collaborations.
In this figure we see excellent agreement between the older quenched
\cite{Gockeler:2004wp,Orginos:2005uy} and $N_f=2+1$ DWF runs and the
$N_f=2$ clover results, with the possible exception of the lightest
clover mass.
This discrepancy may be attributed to a finite size effect $(m_\pi
L=2.78)$, since these effects are expected to enhance $\langle
x\rangle$ at light quark masses \cite{Detmold:2005pt}.

While we see agreement between the DWF and Clover results,
we observe a gap between these results and those coming from the mixed
action approach.
Since the overall pion mass dependence is similar, this suggests that
it is a renormalisation effect; a suggestion further enhanced when we
consider that the results from the mixed approach use
(non-perturbatively improved) perturbative renormalisation
\cite{Hagler:2007xi}, while those from the other approaches use
nonperturbative renormalisation of the operators involved.
Of course, for this issue to be fully resolved, the mixed action
results need to be renormalised nonperturbatively.

The $\chi$QCD collaboration has also started an investigation of
$\langle x\rangle$ using $16^3\times 32$, $N_f=2+1$ Clover
configurations from the CP-PACS/JLQCD collaborations with
$a=0.1219$~fm.
Preliminary results from simulations at $m_\pi\sim 800$~MeV were
presented in \cite{Mankame:2008gt,Deka:2008xr}, with results from
lighter quark masses forthcoming.

RBC/UKQCD also presented results for the nucleon's helicity fraction,
tensor charge and twist-3 matrix element, $d_1$ \cite{Ohta:2008kd}.

\subsection{Operator Product Expansion on the Lattice}
\label{sec:ope}

Moments of the nucleon structure functions can be expanded in the
lattice regularisation as
\be
{\cal M}(q^2) = c^{(2)}(aq) A_2(a) + c^{(4)}\frac{1}{q^2} A_4(a) +
\ldots {\rm higher\ twist}\ ,
\ee
where $q$ is the momentum transfer, $a$ the lattice spacing, $c^{(n)}$
the Wilson coefficients of twist, $n$, and $A_n$ the reduced matrix
elements.

The leading twist matrix elements are nonperturbative quantities and
can be studied on the lattice (an example of which we have just seen
in the previous section).
The corresponding Wilson coefficients, however, are usually calculated
in continuum perturbation theory.
Recently it has been shown that by applying the Operator Product
Expansion to a product of electromagnetic currents between quark
states, it is possible to determine the Wilson coefficients
nonperturbatively \cite{Capitani:1998fe}, allowing for a consistent
treatment of the moments of structure functions.

QCDSF are currently performing a quenched simulation on a $24^3\times
48$ lattice using overlap fermions \cite{Bietenholz:2008fe}, which
have the advantage that undesired operator mixings are suppressed by
chiral symmetry and results are free of $O(a)$ artifacts.
By considering two different momenta, $q=\frac{\pi}{4a}(1,1,1,1)$ and
$\frac{\pi}{3a}(1,1,1,1)$, it was shown that preliminary results for
Wilson coefficients of the 67 operators considered have the correct
Bjorken scaling.
Further improvements will involve using twisted bc's to access smaller
momenta.
With the full data available, a fully nonperturbative and consistent
evaluation of the moments of nucleon structure functions will be
possible.

%%%%%%%%%%%%%%%%%%%%%%%%%%%%%%%%%%%%%%%%%%%%%%%%%%%%%%%%%%%%%%%%%%%
%
\section{Baryon Axial Charges}
\label{sec:axial}
%
%%%%%%%%%%%%%%%%%%%%%%%%%%%%%%%%%%%%%%%%%%%%%%%%%%%%%%%%%%%%%%%%%%%

The axial coupling constant of the nucleon is important as it governs
neutron $\beta$-decay and also provides a quantitative measure of
spontaneous chiral symmetry breaking.
It is also related to the first moment of the helicity dependent quark
distribution functions, $g_A=\Delta u - \Delta d$.
It has been studied theoretically as well as experimentally for many
years and its value, $g_A=1.2695(29)$, is known to very high accuracy.
Hence it is an important quantity to study on the lattice, and since
it is relatively clean to calculate (zero momentum, isovector), it
serves as useful yardstick for lattice simulations of nucleon
structure.

\subsection{$g_A$}
\label{sec:gA}

The axial charge is defined as the value of the isovector axial form
factor at zero momentum transfer and is determined by the
forward matrix element
\be 
\langle p,\,s|A_{\mu}^{u-d}|p,\,s\rangle \, = 2g_A s_\mu \ ,
\ee
where $p$ is the nucleon momentum, and $s_\mu$ is a spin vector with
$s^2=-m_N^2$.

\begin{figure}[t]
   \begin{minipage}{0.48\textwidth}
      \centering
\includegraphics[width=0.9\textwidth]{Figures/mpiL_g_A.eps}
\caption{Scaling of $g_A$ with $m_\pi L$ \cite{Yamazaki:2008py}}
\label{fig:RBCgA}
     \end{minipage}
     \hspace{0.1cm}
    \begin{minipage}{0.48\textwidth}
      \centering
\includegraphics[width=0.94\textwidth]{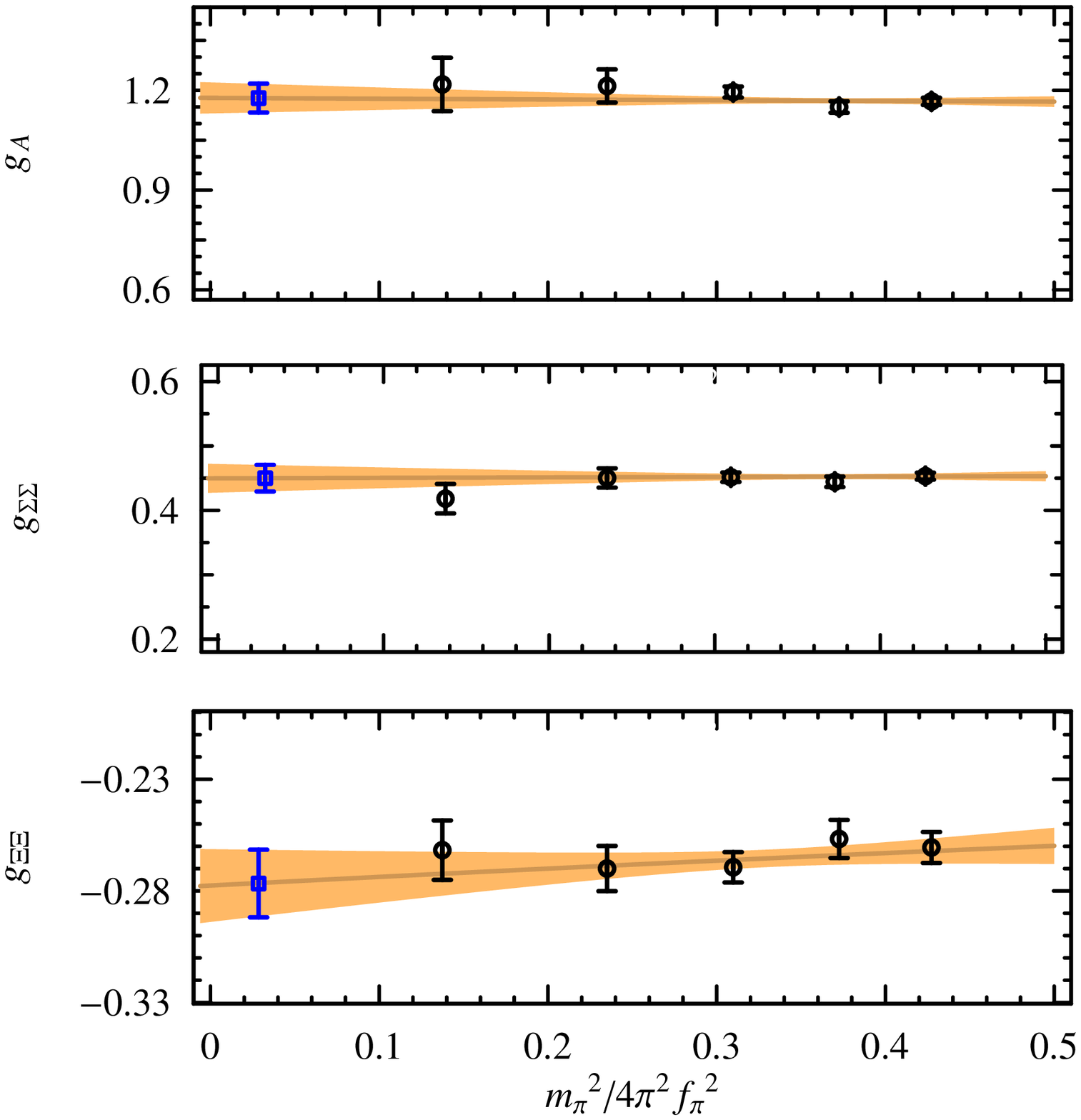}
\vspace*{-4mm}
\caption{$g_A,\, g_{\Sigma\Sigma},\, g_{\Xi\Xi}$ from
  \cite{Lin:2007ap} using a mixed action approach.}
\label{fig:octetgA}
     \end{minipage}
 \end{figure}

$g_A$ has been studied in-depth for many years by the QCDSF
\cite{Khan:2006de} and LHP collaborations \cite{Edwards:2005ym} and
has been shown to suffer from large finite size effects.
The RBC/UKQCD collaborations have recently calculated $g_A$ on their
$N_f=2+1$ DWF configurations \cite{Yamazaki:2008py}, where they
observed the finite size effects to scale exponentially with $m_\pi L$
\cite{Yamazaki:2008py} as seen in Fig.~\ref{fig:RBCgA} for the DWF and
Clover results.

The ETM collaboration have also started simulations to measure $g_A$
on their $N_f=2$ twisted mass lattices, and we saw a status report
\cite{TWgA}.

Finally, LHPC have a new simulated mixed action point at $m_\pi\sim
293$~MeV and have also started to measure $g_A$ on the $N_f=2+1$ DWF
configurations generated by RBC/UKQCD \cite{Bratt:2008uf}.
For the latter, measurements are being performed at three quark masses
and two lattice spacings, but with similar volumes.
Preliminary analysis indicates that results from the two approaches
agree, indicating that effects due to unitarity violation in the mixed
action approach is negligible.

\subsection{Axial Coupling Constants of Octet Baryons}
\label{sec:goctet}

While there has been much work on the (experimentally well-known)
nucleon axial coupling, there has been limited work on the axial
coupling constants of the other octet baryons, which are relatively
poorly known experimentally.
These constants are important since at leading order of SU(3) heavy
baryon ChPT, these coupling constants are linear combinations of the
universal coupling constants $D$ and $F$, which enter the chiral
expansion of every baryonic quantity.

Lin and Orginos \cite{Lin:2007ap} have used DWF valence quarks on an
Asqtad sea with $m_\pi$ ranging between 350 and 750\,MeV and their results for
$g_A,\, g_{\Sigma\Sigma}$ and $g_{\Xi\Xi}$ are shown in
Fig.~\ref{fig:octetgA}.
Fitting all three couplings simultaneously using 
$g_A = D+F+\sum_n C_N^{(n)} x^n$,
$g_{\Xi\Xi} = F-D +\sum_n C_\Xi^{(n)} x^n$,
$g_{\Sigma\Sigma} = F+\sum_n C_\Sigma^{(n)} x^n$,
with $x=(m_K^2 - m_\pi^2)/(4\pi f_\pi^2)$, they find
\be
g_A = 1.18(4)_{\rm stat}(6)_{\rm sys},\ 
g_{\Xi\Xi} = 0.450(21)_{\rm stat}(27)_{\rm sys},\ 
g_{\Sigma\Sigma} = -0.277(15)_{\rm stat}(19)_{\rm sys}\,,
\ee
and $D=0.715(6)(29),\, F=0.453(5)(19)$.
Since there is little known from experiment for $g_{\Xi\Xi}$ and
$g_{\Sigma\Sigma}$, these results serve as a prediction and are
in agreement with findings from ChPT and large-$N_c$.

\subsection{$N^*$ Axial Charges}
\label{sec:gnstar}

In the previous sections, we have seen results for axial couplings of
ground state baryons.
Recently, there has been an attempt to calculate the axial couplings
of the two lowest lying, negative parity nucleon states, the
$N^{*0^-}(1535)$ and $N^{*1^-}(1650)$ \cite{Takahashi:2008fy}.

The authors have used the $16^3\times 32$, $N_f=2$ clover configurations
from the CP-PACS collaboration with $a=0.1555(17)$\,fm and $m_{ps}/m_v
= 0.804(1),\,0.752(1),\, 0.690(1)$.
In order to isolate the two negative parity states, they construct
optimised source/sink operators from a combination of operators.
In order to verify their method, they also calculate $g_A$ of the
nucleon to compare with other determinations.
They are able to see a signal and 
after extrapolating their results linearly in $m_\pi^2$ to the
physical pion mass, they find $g_A^{0-}<0.2,\ g_A^{1-}\approx 0.55$,
which is consistent with the NR quark model.

%%%%%%%%%%%%%%%%%%%%%%%%%%%%%%%%%%%%%%%%%%%%%%%%%%%%%%%%%%%%%%%%%%%
%
\section{Generalised Parton Distributions}
\label{sec:gpds}
%
%%%%%%%%%%%%%%%%%%%%%%%%%%%%%%%%%%%%%%%%%%%%%%%%%%%%%%%%%%%%%%%%%%%

Generalised Parton Distributions (GPDs) have received much attention,
from both theory and experiment, in the past decade since they provide
a solid framework in QCD to relate many different aspects of hadron
physics, including form factors, parton distribution functions, impact
parameter dependent PDFs and spin sum rules.
The importance of these functions has led the QCDSF and LHP
collaborations to perform lattice investigations of their moments
\cite{Hagler:2007xi,Gockeler:2003jfa}, where it is has been shown that
the $q^2$-dependence of the generalised form factors associated with
these moments flatten for increasing moment.
This has the interpretation of a narrowing quark distribution in the
transverse plane of a fast-moving nucleon as $x_q\to 1$.

Here we focus on the insights moments of GPDs provide into the spin
structure of the nucleon.

\subsection{Spin Sum Rules}
\label{sec:ssr}

It is now well known that quark spin carries only $\sim30\%$ of the
total spin of the nucleon, with the remaining $\sim70\%$ coming from
quark orbital angular momentum and glue.
The total spin of the nucleon can be decomposed in terms of the quark
and gluon angular momentum
\be
\frac{1}{2}=\sum_q J_q(\mu^2) + J_g(\mu^2) \ ,
\ee
which is then further decomposed into the quark and gluon spin and
orbital angular momentum contributions
\be
\frac{1}{2}=\sum_q\frac{1}{2}\Delta\Sigma_q + \sum_q L_q + \Delta G +
L_g\ ,
\ee
where $\Delta\Sigma$ and $\Delta G$ are the standard gauge-invariant
quark and gluon spin fractions, while the orbital angular momentum
contributions are defined by $L_q=J_q-\Delta\Sigma/2$ and
$L_g=J_g-\Delta G$.
The relation of total angular momenta, $J_{q,g}$, to the GPDs is due
to Ji\cite{Ji:1996ek} who showed that they can be expressed in terms
of moments of GPDs
\be
J_{q/g}= \frac{1}{2}\bigg[\int dx\,x(H_{q/g}(x,\xi,t) +
E_{q/g}(x,\xi,t)) \bigg]=
\frac{1}{2}\bigg[A_{20}^{q/g}(\Delta^2=0) +
B_{20}^{q/g}(\Delta^2=0)\bigg]\ ,
\ee
where $t=\Delta^2$ and $A,B_{20}$ are matrix elements of the energy
momentum tensor
\be
\langle P'|T^{\mu\nu}|P\rangle = \overline{U}(P')\bigg\{\gamma^\mu
\overline{P}^\nu A_{20}(\Delta^2) +
\frac{i\sigma^{\mu\rho}\Delta_\rho\overline{P}^\nu}{2m_N}
B_{20}(\Delta^2) + \frac{\Delta^\mu\Delta^\nu}{m_N}
C_{20}(\Delta^2)\bigg\} U(P)\ .
\ee
Since $A_{20}^{q,g}(0)=\langle x\rangle^{q,g}$ are simply the quark
and gluon momentum fractions, we have by momentum conservation
$1=\sum_q A_{20}^q(0) + A_{20}^g(0)$, hence we have a sum rule for the 
anomalous gravitomagnetic moments
$0=\sum_q B_{20}^q(0) + B_{20}^g(0)$.
Here we stress that although the sum is scale and scheme independent
and is equal to zero, for the individual $B_{20}^{q,g}$, this is not
necessarily the case.

\begin{figure}[t]
  \begin{tabular}{cc}
  \hspace{-.3cm}  \includegraphics[width=0.48\textwidth]{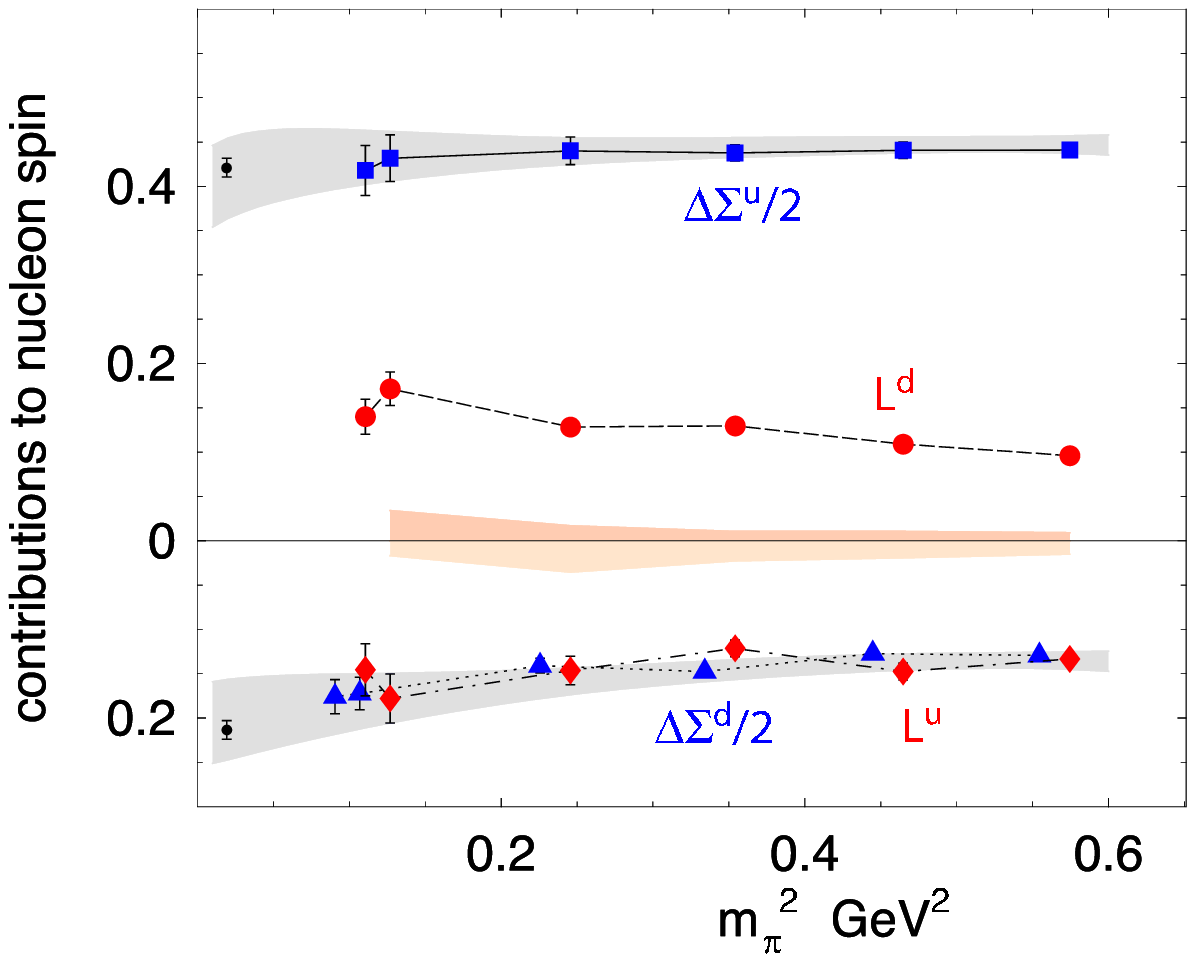} &
   \raisebox{1.5 cm}{ \includegraphics[width=0.48\textwidth]{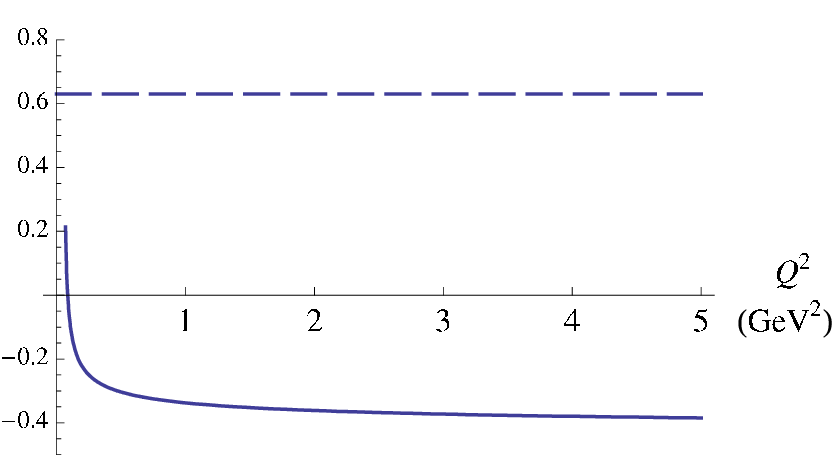}}\\
  \end{tabular}
\caption{LHPC: $\Delta\Sigma^q/2$ and $L^q$ (left) and evolution of
  $L^{u-d}$ with respect to the scale, $Q^2$ (right) \cite{Bratt:2008uf}.}
\label{fig:LHPCspin}
\end{figure}

Most of the work towards a determination of $B_{20}^{q,g}$ has been
done by the LHP \cite{Hagler:2007xi} and QCDSF \cite{Brommel:2007sb}
collaborations.
This year, we have seen an update from LHPC for their simulations
using the mixed action approach (left plot, Fig.~\ref{fig:LHPCspin})
and some preliminary results of a study using the $N_f=2+1$ DWF
configurations from the RBC/UKQCD collaborations \cite{Bratt:2008uf}.
The results in the left of Fig.~\ref{fig:LHPCspin} indicate that the
signs of the spin $(\Delta\Sigma_q)$ and orbital angular momentum
$(L_q)$ contributions are opposite for each quark flavour.
The same behaviour has been observed by QCDSF \cite{Brommel:2007sb}.
The lattice result $L^{u+d} \sim 0$ is in strong disagreement with
relativistic quark models and has led LHPC to search for scale
dependence in $L^q$, as suggested by \cite{Thomas:2008ga}.
As seen in the right of Fig.~\ref{fig:LHPCspin}, they find that
$L^{u-d}$ changes dramatically at small $Q^2$ and in fact changes
sign, which may help to reconcile the lattice and the quark model
results, which generically are valid at a low hadronic scale
$\mu\ll 1$~GeV.
Although as the authors point out, the one-loop evolution used here is
not perhaps not quantitatively reliable below 1~GeV.

A potential improvement in the determination of $B_{20}^q$ from the
lattice is in the extrapolation that is required from the simulated
points at $q^2\ne 0$ to the required point $q^2=0$.
As seen in Sec.~\ref{sec:tbc}, this can be achieved through the use of
twisted boundary conditions, which is currently being explored by the
QCDSF collaboration \cite{Phil-latt}.
This may become particularly important at light quark masses when the
data becomes noisier, and hence the extrapolation is poorly
constrained.

\subsection{Spin Asymmetries}
\label{sec:sasym}

In the past couple of years, lattice calculations of the moments of
GPDs have provided facinating insights into how quarks are spatially
distributed inside the nucleon \cite{Gockeler:2006zu} and pion
\cite{Brommel:2007xd}.
Of particular interest is the strong correlation between the
transverse spin and coordinate degrees of freedom
\cite{Burkardt:2005hp}, providing evidence for a sizeable Boer-Mulders
function, $h_1^\perp(x,k_\perp^2)$ \cite{Boer:1997nt}.

Recently, there has been an attempt to determine on the lattice
(moments of) the Transverse Momentum Dependent PDFs (TMDPDFs), e.g.
$f_1^\perp(x,k_\perp),\,h_1^\perp(x,k_\perp)$, which are important in
semi-inclusive DIS (SIDIS).
In order to obtain information on the dependence of these functions on
the transverse momentum, $k_\perp$, of the quarks inside a hadron, it
is necessary to consider matrix elements
$\langle P|\bar{q}(\ell)\,\Gamma\,{\cal U}\,q(0)|P\rangle$,
where the quark fields are separated by a distance, $\ell$, and
${\cal U}$ is a Wilson line (to infinity and back). 
Of course, this is not possible on the lattice so instead one is
required to consider a path of finite total length $\ell$ separating
the quark and anti-quark in the operator, as illustrated in Fig.~3a of
\cite{Musch:2007ya}.

The matrix element is then obtained from from
\be
\frac{C_{\rm 3pt}(\tau,t_{\rm sink},P,\Gamma)}{C_{\rm 2pt}(t_{\rm
    sink},P)}\overset{0\ll\tau\ll t_{\rm
    sink}}{\longrightarrow}\langle P|\bar{q}(\ell)\,\Gamma\,{\cal
  U}\,q(0)|P\rangle\propto \tilde{A}_i(\ell^2,\ell\cdot P)\ .
\ee
Choosing $\Gamma$ to be $\gamma_\mu$ gives access to
$\tilde{A}_2,\,\tilde{A}_3$, while $\gamma_\mu\gamma_5$ gives
$\tilde{A}_6,\,\tilde{A}_7,\,\tilde{A}_8$.
The $\ell^2$-dependence of these functions is fitted with a double
Gaussian.
Moments of the TMDPDFs are then obtained via a Fourier transform
\be
f^{n=1}_{1,\text{lat}}(\vec k_T) = \int dx\, f_1(x,\vec{k}_\perp)=\int
\frac{d^2 \vec \ell_\perp}{(2 \pi)^2} e^{i\,\vec k_\perp \cdot \vec
  \ell_\perp} 2\tilde{A}_2 ( |\vec \ell_\perp|,0)\ ,
\ee
and similarly for $g_{1T}^{(1){\rm lat}}$ which is obtained from
$\tilde{A}_7$.

\begin{wrapfigure}{r}{0.45\textwidth}
\bc
\vspace*{-17mm}
\includegraphics[width=0.42\textwidth]{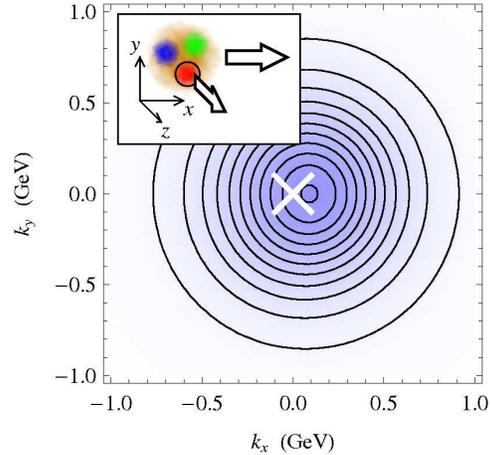}
\vspace*{-2mm}
\caption{TMDPDF for a $u$-quark in a nucleon that is polarised in
  along the $x$-axis}
\label{fig:tmdpdfU}
\vspace*{-10mm}
\ec
\end{wrapfigure}

Information on the correlation between the intrinsic quark transverse
momentum and the transverse polarisation of the nucleon can then be
obtained by considering the combination
\be
\frac{1}{2}\bigg(f_1^{(1){\rm lat}} (k_\perp) +
\frac{k_\perp\cdot S_\perp}{m_N}g_{1T}^{(1){\rm
      lat}} (k_\perp)\bigg)\ ,
\ee
which is shown in Fig.~\ref{fig:tmdpdfU} for a longitudinally
polarised $u$-quark inside a nucleon that is transversly polarised in
along the $x$-axis, $S_\perp=(S_x,0)$.
We clearly see that the distribution is distorted along the $x$-axis.

%%%%%%%%%%%%%%%%%%%%%%%%%%%%%%%%%%%%%%%%%%%%%%%%%%%%%%%%%%%%%%%%%%%
%
\section{Distribution Amplitudes}
\label{sec:da}
%
%%%%%%%%%%%%%%%%%%%%%%%%%%%%%%%%%%%%%%%%%%%%%%%%%%%%%%%%%%%%%%%%%%%

Distribution amplitudes (DAs) describe the momentum-fraction
distribution of partons at zero transverse separation in a particular
Fock state, with a fixed number of constituents.
They are essential for the determination of the hard contributions to
exclusive processes, but being universal hadronic properties, are
process independent.
Hence, they are important for calculations of form factors at large
$Q^2$, B-decays, and can be related to the Bethe-Salpeter wave
function.

DAs are defined as non-local matrix elements on the light cone, e.g.
the leading twist pion DA
\be
\langle 0| \bar d(-z)\gamma_\mu\gamma_5 [-z,z] u(z) |\pi^+(p)\rangle = if_\pi p_\mu 
 \int_{-1}^1 d\xi\, e^{-i\xi p\cdot z}\phi_{\pi}(\xi,\mu^2),\quad
\xi=x-\bar{x}\,.
\ee
Results for moments of the light pseudoscalar meson distribution
amplitudes have been presented by QCDSF \cite{Braun:2006dg} and
UKQCD/RBC \cite{Boyle:2006pw} in the last couple of
years.
Here we will focus on some recent results for vector mesons and the
nucleon.

\subsection{Vector Mesons}
\label{sec:vda}

For spin-1 mesons, there are two DA's,
$\phi^\parallel(\xi),\,\phi^\perp(\xi)$, as opposed to a single
DA for spin-0 mesons.
The lowest moments of $\phi^{\parallel}(\xi)$ are obtained from
the local matrix elements
\begin{eqnarray}
\langle 0 | \bar{q}(0)\gamma_{\{\rho}
\stackrel{\leftrightarrow}{D}_{\mu\}} s(0)| V(p,\lambda) \rangle
&=& m_V f_V p_{\{\rho}\epsilon_{\mu\}}^{(\lambda)}
\langle \xi^1\rangle_V^\parallel\ ,\\
\langle 0 | \bar{q}(0)\gamma_{\{\rho}
\stackrel{\leftrightarrow}{D}_{\mu}
\stackrel{\leftrightarrow}{D}_{\nu\}} q(0)| V(p,\lambda) \rangle
&=& -i m_V f_V p_{\{\rho}p_{\mu}\epsilon_{\nu\}}^{(\lambda)}
\langle \xi^2\rangle_V^\parallel\ ,
\end{eqnarray}
where $m_V$ and $f_V$ are the mass and decay constant, respectively,
of the the vector meson, $V$, and $\epsilon_\mu$ is a polarisation
vector.
The moments, $\langle \xi^n\rangle_V^\parallel$ are extracted by
constructing ratios of lattice two-point functions
\cite{Braun:2007zr,Boyle:2008nj} and the bare lattice results are then
renormalised.

\begin{figure}
\bc
\includegraphics[width=6.5cm]{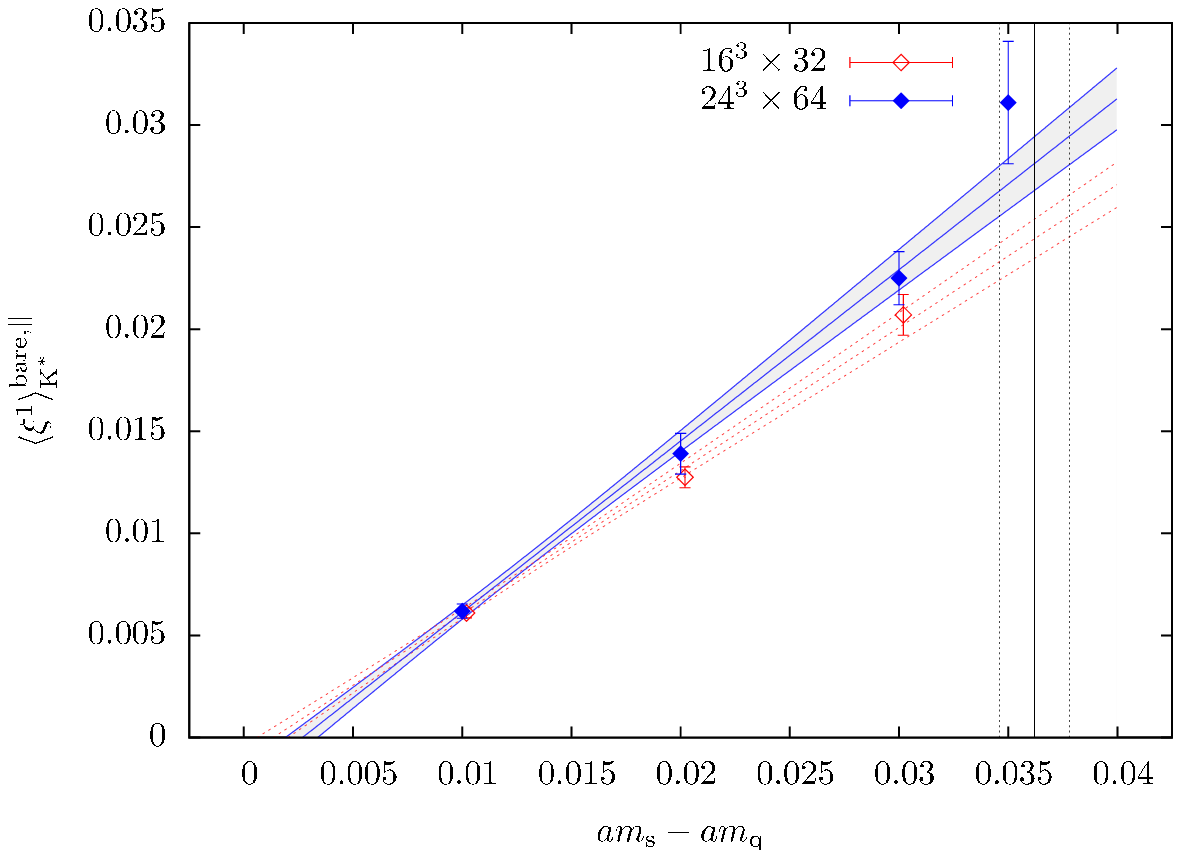}
\hspace*{4mm}
\includegraphics[width=6.5cm]{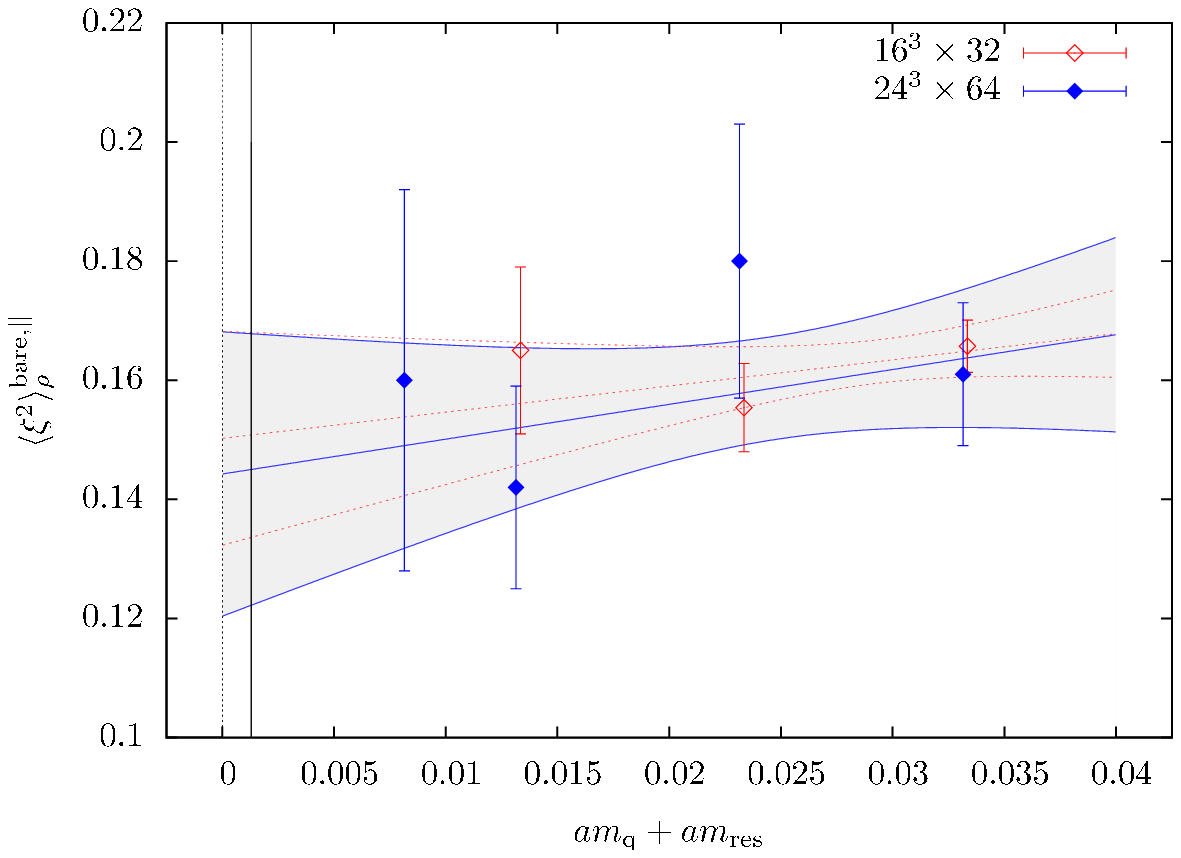}
\caption{$\langle \xi^1\rangle_{K^*}^\parallel$ and $\langle
  \xi^2\rangle_{\rho}^\parallel$ using $N_f=2+1$ DWF \cite{Boyle:2008nj}.}
\label{fig:UKQCDda}
\ec
\end{figure}

In Fig.~\ref{fig:UKQCDda} we see some preliminary results from the
RBC/UKQCD collaborations for $\langle \xi^1\rangle_{K^*}^\parallel$
and $\langle \xi^2\rangle_{\rho}^\parallel$ calculated with $N_f=2+1$
DWF configurations with 4 values of the light quark mass and 2
volumes \cite{Boyle:2008nj}.
The results indicate that there are no clear signs of finite volume
effects.
After renormalising perturbatively (although in \cite{Boyle:2008nj}
they also presented a status report on their nonperturbative
renormalisation programme and the results should be finalised soon) to
the $\overline{\rm MS}$ scheme at $\mu^2=4\,{\rm GeV}^2$, they find
\be
\langle\xi\rangle_{K^*}^{\parallel} \approx
0.0359(17)(22)\quad
\langle\xi^2\rangle_{\rho}^{\parallel}\approx
0.240(36)(12)\quad
\langle\xi^2\rangle_{K^*}^{\parallel} \approx
0.252(17)(12)\ ,
\ee
which compare well with the preliminary results from QCDSF
\cite{Braun:2007zr}
$\langle \xi\rangle_{K^*}^{\parallel}\approx 0.036(3)$,
$\langle \xi\rangle_{K^*}^{\perp}\approx 0.030(2)$.
These results show the $SU(3)_f$-breaking effects in the $K^*$ DAs in
a similar way to that observed for the $K$ DAs in \cite{Braun:2006dg,Boyle:2006pw}.

\subsection{Nucleon}
\label{sec:nda}

For the proton, there are three distribution amplitudes, $V,\, A,\,
T$.
In a similar way to the case of mesons above, their moments
($V^{lmn},\,A^{lmn},\,T^{lnm}$) can be obtained from hadron-to-vacuum
matrix elements of local operators \cite{Kaltenbrunner:2008pb}.
It is useful to construct the combination,
$\phi^{lmn}=\frac{1}{3}(V^{lmn} - A^{lmn} + 2T^{lnm})$.
In the asymptotic limit, $\varphi(x_i,Q^2\to \infty) = 120x_1 x_2 x_3$
and we have
$\phi^{100}=\phi^{010}=\phi^{001}=\frac{1}{3}$,
$\phi^{200}=\phi^{020}=\phi^{002}=\frac{1}{7}$,
$\phi^{110}=\phi^{101}=\phi^{011}=\frac{2}{21}$,
hence it is useful to look for asymmetries, such as
$\phi^{100}-\phi^{010}$.

\begin{figure}
\bc
\includegraphics[width=0.43\textwidth]{Figures/tight_twographs.eps}
\includegraphics[width=0.43\textwidth]{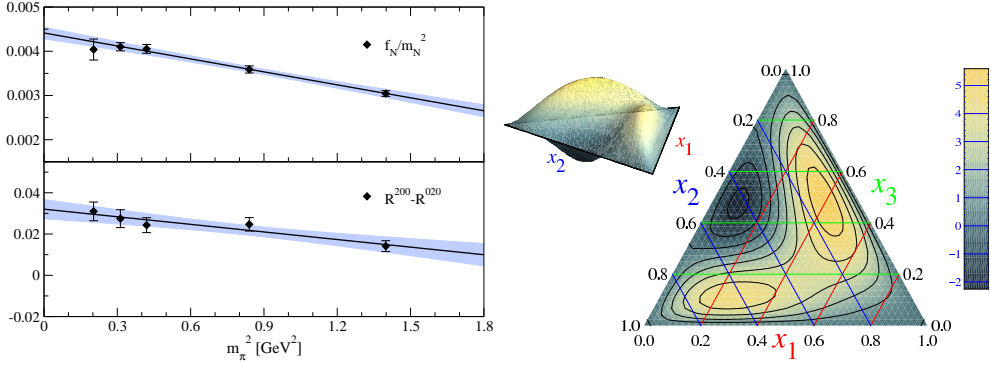}
\caption{Nucleon decay constant, $f_N$, (top/left) and difference
  between ratios of moments (bottom/left) indicating an asymmetry between
  $\phi^{200}$ and $\phi^{020}$. Right: Barycentric contour plot of
  the leading-twist nucleon distribution amplitude at $\mu=2$~GeV.}
\label{fig:ndaratextrap}
\ec
\end{figure}

QCDSF have calculated first two
moments \cite{Gockeler:2008xv} using an improved constrained analysis
which considers ratios of correlators \cite{Gockeler:2008xv} together
with nonperturbative renormalisation of the appropriate 3-quark
operators \cite{Gockeler:2008we}.
By considering the difference between two such ratio, as shown in
Fig.~\ref{fig:ndaratextrap}, the asymmetry is pronounced and increases
as one approaches the chiral limit.

These asymmetries are visualised in Fig.~\ref{fig:ndaratextrap}, where the
lattice moments have been used in a polynomial expansion of the full
nucleon DA.
Here $x_{1,2,3}$ refer to momentum fractions of the three quarks in
the proton and the asymmetries indicate that the $u$-quark with spin
aligned with proton spin has largest momentum fraction ($x_1$).
Interestingly, the asymmetries are less pronounced than for QCD sum
rules \cite{King:1986wi} and other phenomenological determinations
\cite{Bolz:1996sw}.

%%%%%%%%%%%%%%%%%%%%%%%%%%%%%%%%%%%%%%%%%%%%%%%%%%%%%%%%%%%%%%%%%%%
%
\section{Strange Quarks in the Proton}
\label{sec:disc}
%
%%%%%%%%%%%%%%%%%%%%%%%%%%%%%%%%%%%%%%%%%%%%%%%%%%%%%%%%%%%%%%%%%%%

The determination of the strange quark content of the nucleon offers a
unique opportunity to obtain information on the role of hidden flavour
in the structure of the nucleon.
Since the nucleon has no net strangeness, the strangeness
contribution to the total charge of the nucleon must be zero,
i.e. $G_E^s(0)=0$.
However, there is no such simple constraint on either the sign or
magnitude of the strangeness contribution to the magnetic moment,
$G_M^s(0)$.
Additionally, the strangeness charge radius may also be non-zero.
While the latest experimental results \cite{Acha:2006my} suggest that
the strange form factors of the proton are consistent with zero,
forthcoming experiments at JLab and Mainz will further clarify this
picture.
Additionally, the strangeness contribution to the total spin of the
nucleon is poorly determined.
Hence there is an opportunity for lattice simulations to make an
important contribution to the current understanding of the role of
strange quarks in the nucleon.

\subsection{Indirect strangeness}
\label{sec:indirect}

An indirect method for determining the electromagnetic strangeness
form factors has been proposed over the last couple of years by the
Adelaide group \cite{Leinweber:2004tc}.
By combining charge symmetry constraints with chiral extrapolation
techniques, based on finite-range-regularisation \cite{Young:2002ib},
and low-mass quenched-QCD simulations of the individual quark
contributions to the charge radii and magnetic moments of the nucleon
octet, precise estimates of the proton's strange electric charge
radius and magnetic moment were obtained.

\begin{figure}[t]
\vspace*{-2mm}
   \begin{minipage}{0.48\textwidth}
      \centering
\includegraphics[width=0.9\textwidth]{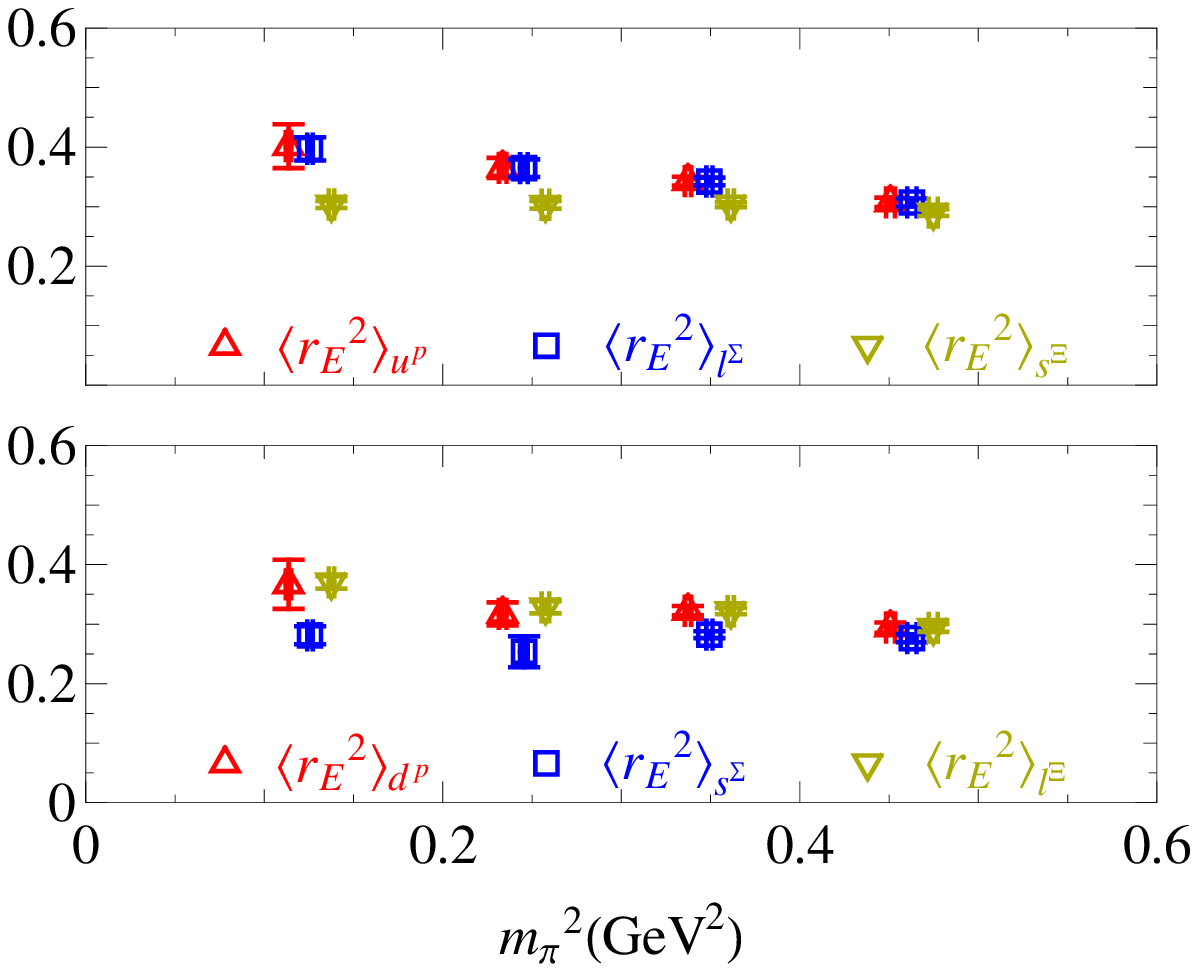}
\caption{Charge radii of the nucleon octet and the contributions from
  the individual quark sectors from a mixed action 
  simulation.}
\label{fig:srad}
     \end{minipage}
     \hspace{0.1cm}
    \begin{minipage}{0.48\textwidth}
      \centering
\includegraphics[angle=-90,width=0.9\textwidth]{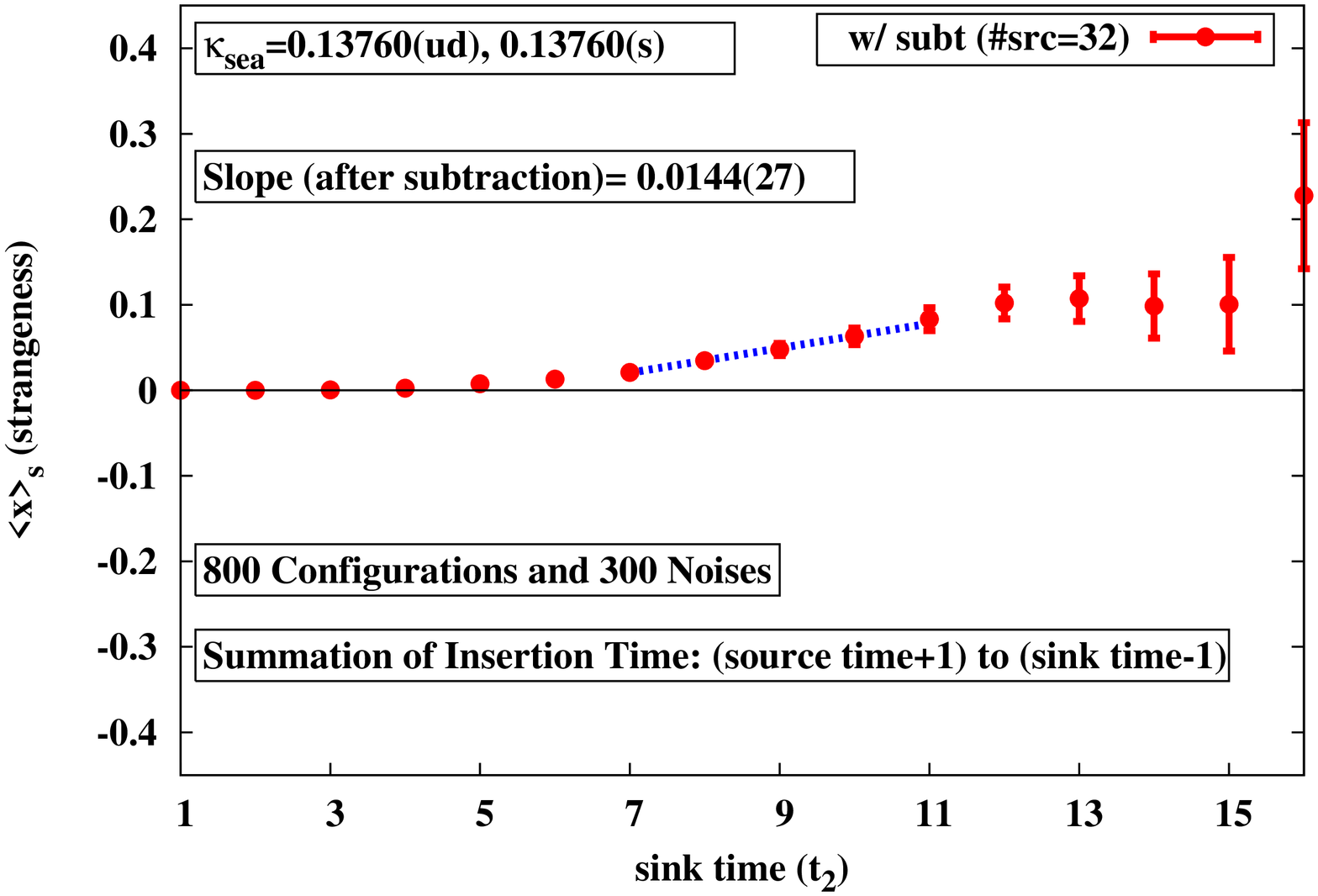}
\caption{Fitting the slope gives a result for $\langle x\rangle_s$
  \cite{Doi:2008hp,Deka:2008xr}.}
\label{fig:Kentuckyxs}
     \end{minipage}
 \end{figure}

Recently, Lin \& Orginos \cite{HW-latt} have followed this procedure
using results from a mixed action simulation (DWF valence on Asqtad
sea) with pion masses in the range $m_\pi=350-750$\,MeV.
Their findings for the individual quark contributions to the charge
radii of the nucleon octet indicate that the contribution from the
heavier strange quark is smaller than those from the light quarks
(Fig~\ref{fig:srad}), in agreement with quenched results
\cite{Boinepalli:2006xd}.

After taking these mixed action results and following the Adelaide
method, Lin \& Orginos find at $Q^2\sim 0.1\,{\rm GeV}^2$
\be
G_M^s = -0.082(8)(25)\,, \quad
G_E^s = -0.00044(1)(130)\ ,
\ee
which is in excellent agreement with earlier findings
\cite{Leinweber:2004tc} and recent JLab experiments
\cite{Acha:2006my}.

\subsection{Direct strangeness and other disconnected}
\label{sec:direct}

Direct lattice calculations of the strangeness content are
computationally demanding and are renowned for suffering from large
statistical noise.
Recent advances in computing power combined with technical
innovations, such as all-to-all propagators \cite{Foley:2005ac}, have
led to a renewed interest in direct determinations of disconnected
quantities, such as strangeness in the nucleon.
This year we have seen the progress being made in this area from
several groups using a variety of different methods.

The $\chi$QCD collaboration \cite{Doi:2008hp,Deka:2008xr} have started a
simulation to determine the gluonic and strange quark momentum
fractions of the nucleon, $\langle x\rangle_g,\ \langle x\rangle_s$,
and the strangeness magnetic moment using $N_f=2+1$ dynamical clover
configurations from the CP-PACS/JLQCD collaborations.
They use $Z(4)$ stochastic noise sources combined with an unbiased
subtraction from the hopping parameter expansion (HPE) \cite{Thron:1997iy},
and multiple (up to 32) sources.
By summing over the operator insertion times, they can then fit to the
slope, as shown in Fig.~\ref{fig:Kentuckyxs}.
Their preliminary findings suggest that the strange-to-light momentum
fraction ratio, 
$\langle x\rangle_{\bar{s}}/(\frac{1}{2}(\langle x\rangle_{\bar{u}} +
  \langle x\rangle_{\bar{d}}))=0.857(40) $, which is slightly larger than the
  CTEQ value, $0.27<r<0.67$.
Additionally, $\langle x\rangle_g$ is studied using the overlap
operator to construct $F_{\mu\nu}$ \cite{Liu:2007hq} in quenched QCD;
since ultraviolet fluctuations are expected to be suppressed due to
the exponentially local nature of the overlap operator.
For $\langle x\rangle_g$, a signal is obtained with $\sim 3\sigma$
accuracy, however renormalisation is required.

Another group that is making substantial progress is the Boston group
\cite{Babich}. 
They are using stochastic sources with maximum dilution with $N_f=2$
Wilson fermions and $m_\pi\approx 400$\,MeV.
By using vacuum-subtracted currents, e.g. $V-\langle V\rangle$, they
fit the three-point function directly using input from the two-point
functions 
to calculate $G_S^s(q^2=0),\ G_A^s(q^2=0) = \Delta s$.
They find that they are able to determine $\Delta s$ with 30\% errors
and their result for $G_M^s$ is consistent with zero.
From the scalar form factor, they obtain the result
$f_{Ts} = \frac{m_s \langle N|\bar s s|N\rangle}{M_N} = 0.48(7)(3)$.

There was an update \cite{Bali} on work outlined in
\cite{Collins:2007mh} to calculate $\Delta s$ and $\langle N|\bar{s}
s|N\rangle$ using various noise reduction techniques such as HPE,
truncation solver method, truncated eigenmode approach and dilution so
that the stochastic source is only defined on a single timeslice.
They find a reduction in the stochastic variance at fixed cost of
around 25-30.
Disconnected loops are currently being calculated using Wilson
propagators on a staggered sea for three valence quark masses and two
sea quark masses, with plans to move on to a full $N_f=2+1$ simulation
in the near future.

\subsubsection{Scalar Form Factor}
\label{sec:scalarff}

An analysis of the chiral behaviour of the scalar radius of the pion,
$\langle r_S^2\rangle$, can lead to a determination of the LEC
$\ell_4$ and it is expected to have an enhanced chiral logarithm as
compared to the vector radius discussed in Sec.~\ref{sec:pionff}.
Hence it is a good place to search for chiral nonanalytic behaviour in
the chiral regime.
However, such a calculation would need to take into account of the
disconnected contribution to the form factor, and as a result it has
received little attention to date.

As mentioned in Sec.~\ref{sec:pionff}, the JLQCD collaboration are
computing all-to-all propagators on their $N_f=2$ overlap
configurations \cite{JLQCD:2008kx}.
These all-to-all propagators allow them to compute the scalar form
factor of the pion, including the contributions coming from
disconnected diagrams.

\begin{figure}[t]
\vspace*{-2mm}
   \begin{minipage}{0.48\textwidth}
      \centering
\includegraphics[clip=true,width=0.9\textwidth]{Figures/r2_s_vs_Mpi2.meas-pole+cubic.r2_v_s_c_v_nnlo.eps}
\vspace*{-6mm}
\caption{Chiral extrapolation of $\langle r^2\rangle_\pi^S$.}
\label{fig:scalarff}
     \end{minipage}
     \hspace{0.1cm}
    \begin{minipage}{0.48\textwidth}
      \centering
\includegraphics[clip=true,angle=-90,width=0.9\textwidth]{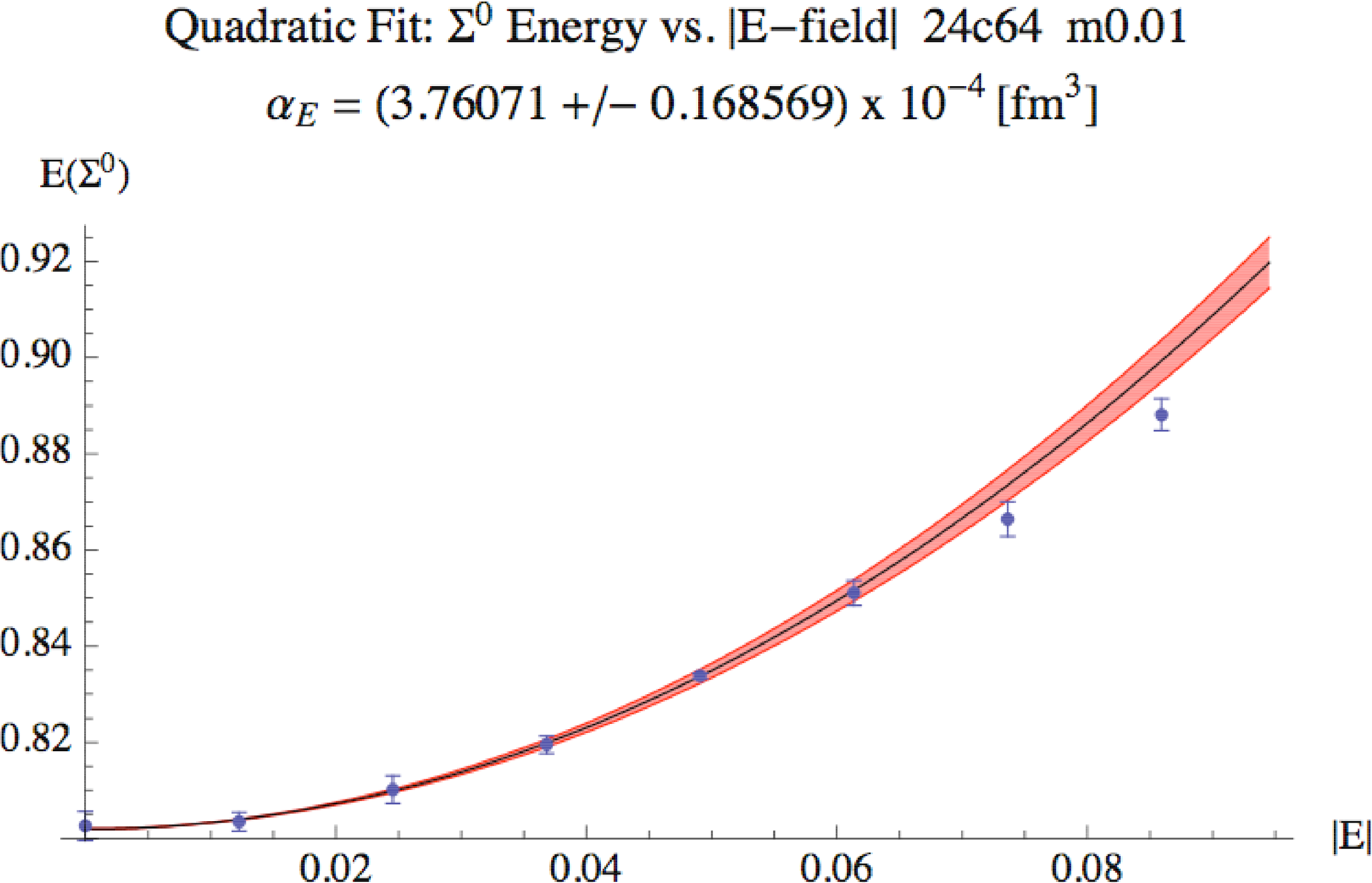}
\caption{Quadratic of the energy of the $\Sigma^0$ as a function of
  electric field.}
\label{fig:epol}
     \end{minipage}
 \end{figure}

From the slope of this form factor, they calculate the scalar radius
of the pion, $\langle r^2\rangle_\pi^S$, which is shown in
Fig.~\ref{fig:scalarff} as a function of $m_\pi^2$.
Also shown in the plot is the result of a combined NNLO ChPT fit to
$\langle r^2\rangle_\pi^S$, $\langle r^2\rangle_\pi^V$ and $c_V$ (see
Sec.~\ref{sec:pionff}), where we can clearly see the predicted chiral
curvature at light quark masses.
It will be interesting to see if this can be confirmed as results
become available at masses below 300~MeV.
After extrapolating to the physical pion mass, they find $\langle
r^2\rangle_\pi^S =0.578(69)(46)\,{\rm fm}^2$, in agreement with
experiment.

%%%%%%%%%%%%%%%%%%%%%%%%%%%%%%%%%%%%%%%%%%%%%%%%%%%%%%%%%%%%%%%%%%%
%
\section{Background Field \& Polarizabilities}
\label{sec:bff}
%
%%%%%%%%%%%%%%%%%%%%%%%%%%%%%%%%%%%%%%%%%%%%%%%%%%%%%%%%%%%%%%%%%%%

All of the results presented in the previous sections have been
obtained using sequential source methods, where a current with a
particular momentum is inserted into one of the quark propagators used
to construct the hadron.
Static quantities are then obtained from a fit to the momentum
transfer, $q^2$, dependence of the resulting form factors.
An alternative method is to consider lattice simulations in the
presence of constant electric, $E$, and magnetic, $B$, fields, and by
studying the energy shifts in a hadron as a function of the field
strength, it is possible to extract not only magnetic moments, but
also electric and magnetic polarisabilities, $\alpha_E$ and $\beta_M$.

On a finite volume, discontinuities can occur as the quark crosses the
boundary of the lattice unless the fields are quantised as
$qa^2B=2\pi n/L$. 
However, this means that for the volumes being used in current lattice
simulations, the applied fields are so large that nonlinearities
can arise (and dominate) in the $B$-dependence of the masses and
possible distortions in the particles themselves.

Aubin {\it et al.} \cite{Aubin:2008hz} studied the finite volume
effects of the magnetic moment of the $\Delta$ baryon.
They perform a test with quenched lattices for two different spatial
volumes ($(1.6\,{\rm fm})^3$ and $(2.4\,{\rm fm})^3$), they implement a
``patch'' to the field by adding the $x$-link modification
$A_\mu(L-1,y,z,t)=-aBLy\delta_{\mu x}$ if $x=L-1$,
resulting a field that is quantised in units of $2\pi/L^2$.
They find that when using unpatched data with $qa^2B<2\pi/L$, large
finite size effects are seen on the small volume.
However, after patching, so long as simulations are performed close to
$2\pi/L^2$, reliable results are obtained, even on the small volume.

Having justified their method, they then proceeded to use $2+1$
flavours of stout-smeared clover fermions on anisotropic lattices with
two spatial volumes and patched magnetic fields.
One light quark propagator ($m_\pi\sim366$~MeV) and one strange quark
propagator to obtain results for the magnetic moments of
$\Delta^{++,+,-,0}$ and $\Omega^-$.
Results for $\Delta^-$ and $\Omega^-$ are consistent with experiment,
with the accuracy of $\Omega^-$ comparable to experiment.

Tiburzi {\it et al.} \cite{Tiburzi:2008pa} calculated $\alpha_E$ for
both neutral and charge hadrons using clover fermions on DWF sea as a
test run, with valence DWF to follow.
They also showed the benefit of patching, although here it is referred
to as ``including transverse links'', by showing that it is possible
to remove ``spikes'' in, e.g. the pion's effective mass, by including
links such as mentioned above.
The electric polarisability, $\alpha_E$, of hadrons are then
determined by examining the quadratic dependence of the energy as a
function of field strength, as in Fig.~\ref{fig:epol}.

For charged hadrons, one must also take into account the sum of the
Born couplings to the particles total charge, leading to a
modification of the time-dependence of the particles two-point
function.
As a result, $\alpha_E$ for charged hadrons are extracted from the
exponential time behaviour of the two-point functions by fitting them
with $exp(-Et - (Q^2{\cal E}^2t^3)/(6M))$.
While errors obtained in this initial study are fairly large, a couple
of particularly interesting results are the ratio
$\alpha_E^{\pi^+}/\alpha_E^{K^+}$, which is found to be in good
agreement with expectations from one-loop chiral perturbation theory,
i.e. it scales as $m_K/m_\pi$, and $\alpha_E$ of $K^{*0}$ and $K^{*+}$
which are found to be negative.
For the proton and neutron, they find $\alpha_E^n=3.6(1.3)$ and
$\alpha_E^p=8.8(5.9)$ $\times 10^{-4}\,{\rm fm}^3$.

Alexandrou \cite{Alexandru:2008sj} showed that an earlier calculation
of $\alpha_E^n$ \cite{Christensen:2004ca} can
be improved by simulating with an exponential background electric
field rather than the linear field used in \cite{Christensen:2004ca}.

Finally, in \cite{Moerschbacher} we saw some preliminary results for
$\beta_M^{p,n}$ on $N_f=2$ clover configurations from the CP-PACS
collaboration with three lattice spacings but constant physical
volume, and quark masses in the range $0.547<m_\pi/m_\rho<0.8$.

%%%%%%%%%%%%%%%%%%%%%%%%%%%%%%%%%%%%%%%%%%%%%%%%%%%%%%%%%%%%%%%%%%%
%
\section{Conclusion \& Outlook}
\label{sec:fin}
%
%%%%%%%%%%%%%%%%%%%%%%%%%%%%%%%%%%%%%%%%%%%%%%%%%%%%%%%%%%%%%%%%%%%

Due to recent computer and algorithmic improvements, lattice 
calculations of hadronic quantities are now becoming available at pion
masses as low as $m_\pi\approx 250$\,MeV, and it is not unreasonable
to expect that soon simulations will be performed close to the
physical pion mass.
However, as we have seen in, e.g. $g_A$, finite size effects (FSE) are
starting to become a serious issue.
As a result, many groups are now planning future simulations on
volumes as big as $(4\,{\rm fm})^3$, in order to minimise these
effects, although corrections from ChPT will still probably need to be
taken into account.

This year we have seen an impressive amount of progress in many
different hadronic quantities, providing fascinating insights into the
structure of hadrons.
From the slope of the electromagnetic form factors, charge radii are
now being computed for hadrons such as $\pi,\,\rho,\, N,\, \Delta$ in
a region where we expect to see dramatic chiral curvature towards the
physical point.
However, as these radii are an indication of the size of a hadron, as
mentioned earlier, FSE need to be considered
carefully.

The $Q^2$ scaling of hadronic form factors is now receiving an
increasing amount of attention. 
In particular, twisted boundary conditions are providing access to
small $Q^2$, but there also is work underway to attempt to probe the
large $Q^2$ region (> 4 GeV$^2$).
The small $Q^2$ region is also an interesting place to study
the Dirac and Sachs electric form factors of the neutron.
The results that are now becoming available at small $Q^2$ are not
only able to help constrain static quantities such as charge radii and
magnetic/quadrupole moments, but also the value of the generalised
form factor $B_{20}(q^2)$, which at $q^2=0$ provides the value of the
anomalous gravitomagnetic moment, which is important in Ji's angular
momentum sum rule.

Lattice calculations of moments of generalised parton distributions
are providing insights into the different quark contributions to the
nucleon's spin and angular momentum, and current results indicate 
$J_u\approx 46\%,\ J_d\approx 0,\ L_{u+d}\approx 0$.
These moments are also providing evidence for non-trivial transverse spin
densities in the pion and nucleon.

Simulations with zero momentum transfer lead to moments of ordinary
parton distribution functions, and include phenomenologically
interesting quantities such as $g_A$ and $\langle x\rangle$.
Here, FSE appear to playing an important role in the
extraction of these quantities, especially for $g_A$ where we have
seen FSE lowering the lattice results.
While there appears to be a slight tension between the renormalisation
of some of the lattice results for $\langle x\rangle_{u-d}$, the
overall pattern seems to indicate that we may now at last be entering
the region where the results may start to ``bend down'' towards the
phenomenological value.
Although once again, FSE are predicted to become an issue close to the
physical pion mass, so care will need to be taken to ensure this
encouraging behaviour continues.

Following the recent success of lattice calculations of the moments of
the light pseudoscalar meson distribution amplitudes (DAs), there are now
results becoming available for moments of vector meson and proton DAs.
Results for the proton are providing evidence that asymmetries exist
in the way the momentum of the nucleon is distributed amongst its
constituent quarks, with the $u$-quark with its spin aligned to that
of the proton carrying the most momentum.
The results also indicate that the symmetries are less pronounced than
in QCD sum-rules.

While it is important to push these more ``standard'' hadronic
measurements as far as we can with the new sets of dynamical
configurations that are becoming available, it is also important to
develop new ideas and techniques.
This year we saw a number of innovative methods for accessing less
well known quantities.

By considering matrix elements of operators where the quark fields are
spatially separated, moments of Transverse Momentum Dependent PDFs
have been computed.
From these moments, it has been seen that densities of longitudinally
polarised quarks in a transversely polarised nucleon are deformed.

To date, lattice calculations of hadronic quantities have neglected
the contributions coming from disconnected diagrams, since these are
notoriously difficult to compute.
Recently, however, there has been a renewed interest in determining
these disconnected contributions to investigate the strangeness and
gluonic content of the nucleon and, in particular, their contributions
to nucleon spin.
Through the use of all-to-all propagators and various noise
reduction techniques, it may now be possible to calculate some of
these contributions with as small as 10\% errors.

Although background field methods have been around for a long time,
they have only recently received a lot of interest, since
traditionally the electromagnetic fields induced on currently sized
lattices were too large.
However recent developments show that it is now possible to consider
fields that are a factor of $L$ smaller.
As a result, magnetic moments and polarisabilities can now be
extracted from these simulations with much more confidence.
Additionally, it has recently been shown that it is possible to
extract the electric polarisabilities of charged hadrons from a
lattice simulation, and preliminary results are promising.

In summary, lattice simulations of hadronic observables have received
a surge of interest over the past few years, such that we are now not
only in a position to confirm experimental findings from a
first-principles calculation, but also to provide predictions for, and
in some cases to guide, future experimental programmes.

%%%%%%%%%%%%%%%%%%%%%%%%%%%%%%%%%%%%%%%%%%%%%%%%%%%%%%%%%%%%%%%%%%%
%
\section*{Acknowledgements}
%
%%%%%%%%%%%%%%%%%%%%%%%%%%%%%%%%%%%%%%%%%%%%%%%%%%%%%%%%%%%%%%%%%%%

It is a pleasure to thank C.~Alexandrou, R.~Babich, D.~Br\"ommel,
W.~Detmold, T.~Doi, Ph.~H\"agler, M.~G\"urtler, A.~J\"uttner,
T.~Kaneko, H.W.~Lin, P.~Moran, B.~Musch, J.~Negele, D.~Pleiter,
G.~Schierholz, S.~Simula, B.~Tiburzi, N.~Warkentin and T.~Yamazaki for
useful discussions and for providing me with many results and figures.
I would also like to thank Ph.~H\"agler and R.~Horsley for careful
proofreading.
This work is supported through the UK's {\it STFC Advanced Fellowship
  Programme} under contract number PP/F009658/1.

%%%%%%%%%%%%%%%%%%%%%%%%%%%%%%%%%%%%%%%%%%%%%%%%%%%%%%%%%%%%%%%%%%%
%

\end{document}